\documentclass[review,authoryear]{elsarticle}
\makeatletter
\def\ps@pprintTitle{%
 \let\@oddhead\@empty
 \let\@evenhead\@empty
 \def\@oddfoot{\centerline{\thepage}}%
 \let\@evenfoot\@oddfoot}
\makeatother
\usepackage{epsfig,graphicx,setspace,multirow}
\usepackage{amsmath,amsthm,amssymb,enumerate,color}
\usepackage[noend]{algorithmic}
\usepackage{setspace,paralist,algorithm}
\usepackage{bm,verbatim,subcaption}
\usepackage{color}
\usepackage{lscape}
\usepackage{xcolor,colortbl}
\usepackage[authoryear]{natbib}
\usepackage{stackrel}

\usepackage{threeparttable}
\usepackage{CJK}
\usepackage[titletoc]{appendix}
\usepackage{titlesec}
\usepackage{slashbox}
\usepackage{accents}

\def\bg{\begin{figure}[tpbh]\begin{center}}
\def\eg{\end{center}\end{figure}}

\numberwithin{equation}{section} 

\long\def\symbolfootnote[#1]#2{\begingroup\def\thefootnote{\fnsymbol{footnote}}
\footnote[#1]{#2}\endgroup}

\newtheorem{thm}{Theorem}[section]\newtheorem{rem}{Remark}[section]
\newtheorem{exam}{Example}[section]
\newtheorem{prop}{Proposition}[section]

\newcommand{\Lower}[1]{\smash{\lower 1.5ex\hbox{#1}}}

\newcommand{\beqn}{\begin{eqnarray}}\newcommand{\eeqn}{\end{eqnarray}}
\newcommand{\beq}{\begin{equation}}\newcommand{\eeq}{\end{equation}}
\def\beqs{\begin{equation*}}\def\eeqs{\end{equation*}}

\def\bg{\mathbf{g}}

\def\.{$.$}

\def\boxit#1{\vbox{\hrule\hbox{\vrule\kern6pt\vbox{\kern6pt#1\kern6pt}\kern6pt\vrule}\hrule}}

\setcitestyle{number}

\begin{document}
\begin{frontmatter}

\title{Maximum Likelihood Estimation of a Semiparametric Two-component Mixture Model using Log-concave Approximation}

\author[address1]{Yangmei Zhou}
\ead{yzhou037@ucr.edu}
\author[address1]{Weixin Yao\corref{mycorrespondingauthor}}
\ead{weixin.yao@ucr.edu}
\cortext[mycorrespondingauthor]{Yao's research is supported by NSF grant DMS-1461677
and Department of Energy with the award DE-EE0007328.}


\address[address1]{Department of Statistics, University of California, Riverside, California 92521, U.S.A.}

\begin{abstract}
Motivated by studies in biological sciences to detect differentially expressed genes,  a semiparametric two-component mixture model with one known component is being studied in this paper. Assuming the density of the unknown component to be log-concave, which contains a very broad family of densities, we develop a semiparametric maximum likelihood estimator and propose an EM algorithm to compute it. Our new estimation method finds the mixing proportions and the distribution of the unknown component simultaneously. We establish the identifiability of the proposed semiparametric mixture model and prove the existence and consistency of the proposed estimators. We further compare our estimator with several existing estimators through simulation studies and apply our method to two real data sets from biological sciences and astronomy.
\end{abstract}


\end{frontmatter}


\section{Introduction}
In this paper, we consider the following two-component mixture model,
\begin{equation}\label{model}
g(x)=(1-p)f_0(x)+pf(x),
\end{equation}
where the probability density function (pdf) $f_0(x)$ is known, whereas the mixing proportion $p\in[0,1]$ and the pdf $f$ are unknown. 
Model (\ref{model}) is motivated by studies in the biological sciences to cluster differentially expressed genes in microarray data \cite{bordes2006semiparametric}. Typically we build a test statistic, say $T_i$, for each gene $i$. Under the null hypothesis, which presumes no difference in expression levels under two or more conditions, $T_i$ is assumed to have a known distribution (in general Student's or Fisher). Under the alternative hypothesis, the distribution is unknown. Thus, the distribution of the test statistic is modelled by (\ref{model}) where $p$ is the proportion of non-null statistics. The estimation of $p$ and the pdf $f$ can tell us the probability $P_i$ that gene $i$ is differentially expressed given $T_i=t_i$:
   $$P_i=\frac{pf(t_i)}{(1-p)f_0(t_i)+pf(t_i)}.$$ \cite{bordes2006semiparametric} considered model (\ref{model}) where $f$ is assumed to be symmetric. They obtained some identifiability results under moment and symmetry conditions,  and proposed to estimate this model under the symmetry of $f$. In addition, they proved the consistency of their estimator under mild conditions.

\cite{song2010estimating} considered another special case,
$$g(x)=(1-p)\phi_\sigma(x)+pf(x)$$
where $f_0=\phi_\sigma$ is a normal density with mean $0$ and unknown standard deviation $\sigma$. This model was inspired by sequential clustering \citep{song2009sequential}, which finds candidates for centers of clusters first,  then carries out a local search to find the objects that belong to those clusters, and finally selects the best cluster. \cite{song2010estimating} proposed an EM-type estimator and a maximizing $\pi$-type estimator for their model which can be easily extended to models where $f_0$ is not normal.

A slightly different model is considered by \cite{xiang2014minimum},
$$g(x)=(1-p)f_0(x; \xi)+pf(x-\mu),$$
where $\xi$ is a possibly unknown parameter, and $\mu$ is a non-null location parameter for $f$. They proposed a new effective estimator based on the minimum profile Hellinger distance (MPHD). They established the existence and uniqueness of their estimator and also proved its consistency under some regularity conditions. Their method actually does not require $f$ to be symmetric and thus can be applied to a more general model. For some other alternative estimators, see, for example \cite{patra2016estimation,ma2015flexible}.


In this paper, we propose to estimate \eqref{model} using a new approach by imposing a log-concave assumption on $f$, i.e. $\text{log}(f)\in\Phi^1$; here $\Phi^d$ denotes the family of concave functions $\phi$ on $R^d$ which are upper semicontinuous and coercive in the sense that $\phi(x)\rightarrow -\infty, \ \text{as} \ ||x||\rightarrow \infty$. Note that $\text{log}(f)$ needs to be coercive in order for $f$ to be a density function.  Many common parametric families of distributions belong to the family of log-concave densities, for example, normal distribution, exponential distribution, logistic distribution, etc. We propose to estimate the new model by maximizing a semiparametric mixture likelihood. Compared to the kernel density estimation of $f$ used by many existing methods  \citep{bordes2006semiparametric,xiang2014minimum,ma2015flexible}, the new method does not require the choice of one or more bandwidths (\cite{samworth2017recent}).  We establish the identifiability of the proposed semiparametric mixture model and prove the existence and consistency of the proposed estimators. We further compare our estimator with several existing estimators through simulation studies and apply our method to two real data sets from biological sciences and astronomy.



The rest of the paper is organized as follows. In Section \ref{identifiability} we discuss some identifiability issues for model (\ref{model}). Section \ref{mle} introduces our maximum likelihood estimator and a detailed EM type algorithm is provided.  Existence and consistency properties of our estimator are established with detailed proofs given in the Appendix (Section \ref{appendix}). Section \ref{simulation} demonstrates the finite sample performance of our proposed estimator by comparing with many other existing algorithms. Two real data applications are given in Section \ref{realdata}. Section \ref{discussion} gives a brief discussion.\\

\section{Identifiability}\label{identifiability}

Note that the model (\ref{model}) is non-identifiable without any constraint on the density $f$, see e.g.,  \cite{bordes2006semiparametric}, and \cite{patra2016estimation}. However a parametric model for $f$ might create biased or even misleading statistical inference when the model assumption is incorrect. In this paper, we assume $f(x)$ to be log-concave, i.e. $f(x)=e^{\phi(x)}$, where $\phi(x)$ is a concave function. 
Log-concave densities attracted lots of attention in the recent years since it is very flexible and can be estimated by nonparametric maximum likelihood estimator without requiring the choice of any tuning parameter. For more details, see \cite{cule2010maximum}, \cite{dumbgen2009maximum},  \cite{walther2009inference}, \cite{dumbgen2011approximation} and the review of the recent progress in log-concave density estimation by  \cite{samworth2017recent}.


\begin{prop}\label{identifiability1}
\label{patra}
Assume $f_0>0$ around a neighborhood of $a$, then model (\ref{model}) is identifiable if,
$$\lim_{x\rightarrow a^+}\frac{f(x)}{f_0(x)}=0  \ \text{or} \
\lim_{x\rightarrow a^-}\frac{f(x)}{f_0(x)}=0. $$
\end{prop}

\begin{rem}
Proposition \ref{patra} also holds if $a=\pm\infty$, and this result is much more general (require much weaker condition) than the result of Proposition 3(i) of \cite{bordes2006semiparametric}.
\end{rem}

\begin{rem}\label{support}
Proposition \ref{patra} guarantees that model (\ref{model}) is identifiable if the support of $f$ is strictly contained in the support of $f_0$ and the two supports have different Legesgue measure.
\end{rem}

Now, for a log-concave density, it is easy to derive the following more specific result,

\begin{prop}
	\label{identifiability2}
Assume $f_0>0$ and $\text{log}(f)\in\Phi^1$. Model (\ref{model}) is identifiable if $\textup{log}f_0(x)=O(x^k)$, for some $0<k<1$,  as  $x\rightarrow +\infty$ or $x\rightarrow -\infty$.
\end{prop}


\begin{exam}\label{students t}
If $f_0(x)$ is the density of a $t$ distribution with $\nu$ degrees of freedom, then model (\ref{model}) is identifiable.
\end{exam}

\begin{proof}
Since $f_0(x)=\frac{\Gamma(\frac{\nu+1}{2})}{\sqrt{\nu\pi}\Gamma(\frac{\nu}{2})}(1+\frac{x^2}{\nu})^{-\frac{\nu+1}{2}}$, we have,
$$\text{log}(f_0(x))=\text{log}(\Gamma(\frac{\nu+1}{2}))-\frac{1}{2}\text{log}(\nu\pi)-\text{log}(\Gamma(\frac{\nu}{2}))-\frac{\nu+1}{2}\text{log}(1+\frac{x^2}{\nu}).$$
Thus for any $0<k<1$, $\text{log}(f_0(x))/x^k\rightarrow 0$, as  $x\rightarrow +\infty$. And by Proposition \ref{identifiability2}, we can conclude that model (\ref{model}) is identifiable when $\text{log}(f)\in\Phi^1$.
\end{proof}

\begin{rem}
Similarly, one can check that when $f_0$ is the pdf of an $F$ distribution, log-normal distribution, or Pareto distribution, then model (\ref{model}) is identifiable under the condition that $\text{log}(f)\in\Phi^1$.
\end{rem}


In general, we need $\phi(x)-\text{log}f_0(x)\rightarrow -\infty$ as $x\rightarrow +\infty$ or $x\rightarrow -\infty$ to make model (\ref{model}) identifiable. Since $f_0(x)$ is known, it is not difficult to give some sufficient conditions with respect to different $f_0$'s. Here we give two examples of easy to apply conditions.

\begin{exam}\label{normal}
Suppose $f_0(x)$ is the density of a normal distribution with mean $\mu$ and variance $\sigma^2$, then model (\ref{model}) is identifiable if $\lim_{x\rightarrow +\infty}\frac{\phi(x)}{x^2}<-\frac{1}{2\sigma^2}$, or $\lim_{x\rightarrow -\infty}\frac{\phi(x)}{x^2}<-\frac{1}{2\sigma^2}$, or the condition of Remark \ref{support} holds.
\end{exam}

\begin{proof}
Suppose $\lim_{x\rightarrow +\infty}\frac{\phi(x)}{x^2}<-\frac{1}{2\sigma^2}$, or $\lim_{x\rightarrow -\infty}\frac{\phi(x)}{x^2}<-\frac{1}{2\sigma^2}$. Since
\begin{eqnarray}
\phi(x)-\text{log}f_0(x)&=&\phi(x)+\text{log}(\sqrt{2\pi}\sigma)+\frac{1}{2\sigma^2}(x-\mu)^2\nonumber\\
&=& x^2(\frac{\phi(x)}{x^2}+\frac{1}{x^2}{\text{log}(\sqrt{2\pi}\sigma)+\frac{1}{2\sigma^2}(1-\frac{\mu}{x})^2})\nonumber\\
&\rightarrow & -\infty, \ \text{as} \  x\rightarrow +\infty \ \text{or} \ x\rightarrow -\infty.\nonumber
\end{eqnarray}

Hence $\displaystyle\frac{f(x)}{f_0(x)}\rightarrow 0$ as $x\rightarrow +\infty$ or $x\rightarrow -\infty$, and Proposition \ref{patra} asserts the identifiability of model (1).
\end{proof}

\begin{rem}
Under the constraints set by Example \ref{normal}, Model 1, 4, and 5 from Section \ref{simulation} are identifiable.
\end{rem}

\begin{exam}\label{exp}
Suppose $f_0(x)$ is the density of an exponential distritution with rate $\lambda$, then model (\ref{model}) is identifiable if $\lim_{x\rightarrow +\infty}\frac{\phi(x)}{x}<-\lambda$, or the condition of Remark \ref{support} holds.
\end{exam}
\begin{proof}
Suppose $\lim_{x\rightarrow +\infty}\frac{\phi(x)}{x}<-\lambda$. Since,
\begin{eqnarray}
  \phi(x)-\text{log}f_0(x) &=& \phi(x)-\text{log}\lambda+\lambda x \nonumber\\
  & = & x(\frac{\phi(x)}{x}-\frac{\text{log}\lambda}{x}+\lambda)\nonumber\\
  &\rightarrow & -\infty, \ \text{as} \ x\rightarrow +\infty.\nonumber
\end{eqnarray}
Hence $\displaystyle\frac{f(x)}{f_0(x)}\rightarrow 0$ as $x\rightarrow +\infty$, and again Proposition \ref{patra} ensures the identifiability of model (\ref{model}).
\end{proof}

\begin{rem}
Under the constraints set by Example \ref{exp}, Model 3 from Section \ref{simulation} is identifiable.
\end{rem}\hfill

\section{Maximum Likelihood Estimation}\label{mle}
Suppose we have a random sample of $n$ i.i.d. observations $X_1, X_2,\cdots, X_n$ from the density $g(x)=(1-p)f_0(x)+pf(x)$, $p\in[0,1]$,  and $f=e^\phi$ is a log-concave density, i.e., $\phi\in\Phi^1$. Then with the empirical distribution $Q_n=\displaystyle\frac{1}{n}\sum_{i=1}^n\delta_{X_i}$ , where $\delta_{X_i}$ is the degenerate distribution function at $\{X_i\}$, the log likelihood of our random sample can be written as,
$$L(p,\phi,Q_n)=n\int \text{log}(g)dQ_n=\sum_{i=1}^n\text{log}((1-p)f_0(X_i)+pe^{\phi(X_i)}),$$
subject to the condition that $\int e^{\phi(x)}dx=1$. A natural approach to estimate $p$ and $\phi$ is to find the maximizer of $L(p,\phi,Q_n)$.\\

\subsection{Algorithm}

 We propose to estimate $p$ and $f$ by maximizing $L(p,\phi,Q_n)$, using the EM algorithm that consists of iterating an E step and M step until convergence:

\hspace{-0.6cm}{\bf``E step"}: Given $p^{(k)}$ and $f^{(k)}$,
\begin{equation}
\omega_i^{(k+1)}=\frac{(1-p^{(k)})f_0(x_i)}{(1-p^{(k)})f_0(x_i)+p^{(k)}f^{(k)}(x_i)},\nonumber
\end{equation}

\hspace{-0.6cm}{\bf``M step"}:
\begin{align}
&p^{(k+1)}=\frac{1}{n}\sum_{i=1}^n(1-\omega_i^{(k+1)}),\nonumber\\
&\phi^{(k+1)}=\stackrel[\phi\in \Phi^1, \ \int e^{\phi(x)}dx=1]{}{\text{argmax}}\sum_{i=1}^n(1-\omega_i^{(k+1)})\phi(x_i),\nonumber\\
&f^{(k+1)}=e^{\phi^{(k+1)}}.\nonumber
\end{align}

We find $\phi^{(k+1)}$ using an active set algorithm, which is described in  \cite{dumbgen2007active} and implemened in the R package \emph{logcondens} by  \cite{rufibach2010logcondens}. Through out this paper, we use ``EM\_logconcave" to represent our method. The following result establishes the monotone properties of EM\_logconcave algorithm.

\begin{prop}
Let $l^{(k)}=\stackrel[i=1]{n}{\sum}\textup{log}((1-p^{(k)})f_0(x_i)+p^{(k)}e^{\phi^{(k)}(x_i)})$, then
$$l^{(k+1)}\geq l^{(k)},$$
for any $k\geq 0$.
\end{prop}\hfill

\subsection{Theoretical Properties}

In general, for any distribution $Q$ on $R^d$, we define,
\begin{eqnarray}
  L(p,\phi,Q)&=&\int \text{log}((1-p)f_0+pe^\phi)dQ, \nonumber
\end{eqnarray}

For the existence of a maximizer of $L(p,\phi,Q)$, we follow the approach of  \cite{dumbgen2011approximation}. We define the convex support of $Q$ as,
$$\text{csupp}(Q)=\bigcap\{C:C\subseteq R^d \ \text{closed and convex}, \ Q(C)=1\}.$$

\begin{thm}\label{existence}
For fixed $f_0$, assume $\textup{supp}\{f_0\}\subseteq \textup{csupp}(Q)$, and there exist some integer $k\geq 1$, such that,
$$\int ||x||^kQ(dx)< \infty \ \ \textup{and} \ \  \textup{interior}(\textup{csupp}(Q))\neq  \emptyset.$$

Let $\tilde{\Phi}^d=\{\phi\in \Phi^d: \int e^{\phi(x)}dx=1\ \text{and}\ f_0(x)\leq m(x)e^{\phi(x)}\ \text{for \ some \ }m(x)=c_0e^{c_1||x||^k}, c_0\geq 0, c_1\geq 0\}$, then
$$  L(Q)=\stackrel[{p\in[0,1],\ \phi\in\tilde{\Phi}^d}]{}{\text{sup}}L(p,\phi,Q)$$
is real. In that case, there exists,
$$(p_0,\phi_0)\in\stackrel[{p\in[0,1],\phi\in\tilde{\Phi}^d}]{}{\textup{argmax}}L(p,\phi,Q).$$
Moreover,
$$\textup{interior}(\textup{csupp}(Q))\subseteq \textup{dom}(\phi_0)=\{x\in R^d: \phi_0(x)>-\infty\}\subseteq \textup{csupp}(Q).$$
\end{thm}
The proof of Theorem \ref{existence} is given in the Appendix (Section \ref{appendix}).


\begin{exam}
Assume $Q$ represents the distribution of Model 1: $g(x)=(1-p)*N(\mu=0,\sigma=2)+p*N(\mu=3,\sigma=1)$, hence $f_0$ represents the pdf of $N(\mu=0,\sigma=2)$ distribution. For any integer $k\geq 2$, $\tilde{\Phi}^1$ contains all the pdfs of normal distribution, logistic distribution, and Laplace distribution, etc. And Theorem \ref{existence} implies that the maximum of $L(p,\phi,Q)$ exists over $p\in [0,1]$ and $\phi\in\tilde{\Phi}^1$.
\end{exam}

In general, the maximizer of $L(p,\phi,Q)$ is not unique. But if $Q$ has density $g_0(x)=(1-p_0)f_0(x)+p_0e^{\phi_0(x)}$, where $g_0(x)$ is identifiable, then $L(p_0,\phi_0,Q)=\int \text{log}(g_0(x))g_0(x)dx$, and this $(p_0,\phi_0)$ is the unique maximizer.  This is because as noted by  \cite{dumbgen2011approximation}, if we have $(p_1,\phi_1)$, such that $L(Q)=L(p_0,\phi_0,Q)=L(p_1,\phi_1,Q)$, let $g_1(x)=(1-p_1)f_0(x)+p_1e^{\phi_1(x)}$, then
$$\int \text{log}(g_0(x)/g_1(x))g_0(x)dx=0.$$
Note the above integral is exactly the Kullback-Leibler divergence which is positive and equals $0$ iff $g_0=g_1$ almost everywhere. Thus $(p_0,\phi_0)=(p_1,\phi_1)$ except that $\phi_0$ and $\phi_1$ may differ on a set of Lebesgue measure zero.

Next we establish the consistency of our maximum likelihood estimator. First, we introduce some notations,
\begin{eqnarray}
  \mathcal{Q}^k &=& \{Q \ \text{on} \  R^d: \int ||x||^kQ(dx)<\infty\}, \nonumber\\
  \mathcal{Q}_0 &=& \{Q \ \text{on} \  R^d: \text{interior}(\text{csupp}(Q))\neq\emptyset\}.  \nonumber
\end{eqnarray}

In what follows, we consider the convergence of distributions under Mallows' distance $D_1$(\cite{mallows1972note}). Specifically, for two distributions $Q$, $Q'\in\mathcal{Q}^k$,
$$D_k(Q,Q')=\stackrel[{\stackrel[X\sim Q,\ X'\sim Q']{}{X,\ X'}}]{}{\text{inf}}\{E||X-X'||^k\}^{1/k}.$$
It is known that $\stackrel[n\rightarrow \infty]{}{\text{lim}}D_k(Q_n, Q)\rightarrow 0$ is equivalent to $Q_n\rightarrow_w Q$ and  $\int ||x||^k Q_n(dx)\rightarrow \int ||x||^kQ(dx)$(\cite{bickel1981some}; \cite{mallows1972note}). Here $Q_n\rightarrow_w Q$ means weak convergence, or convergence in distribution.

Now we are ready to state our consistency theorem.
\begin{thm}\label{consistency}
Assume, (a). $\textup{supp}\{f_0\}\subseteq \textup{csupp}(Q)$; (b). for some fixed integer $k\geq 1$, the unknown density $f$ satisfies the following condition: $\exists m(x)=c_0e^{c_1||x||^k}$, where $c_i\geq 0$, $i=0,1$, such that,
 $f_0(x)\leq m(x)f(x)=m(x)e^{\phi(x)}$. Let $\{Q_n\}$ be a sequence of distributions in $\mathcal{Q}_0\bigcap\mathcal{Q}^k$ such that $\stackrel[n\rightarrow \infty]{}{\textup{lim}}D_k(Q_n,Q)=0$ for some $Q\in\mathcal{Q}_0\bigcap \mathcal{Q}^k$. Suppose $f_0$ is upper semi-continuous and $\displaystyle\frac{\text{log}(f_0)}{1+||x||}$ is bounded. Then
$$\stackrel[n\rightarrow\infty]{}{\textup{lim}}L(Q_n)=L(Q).$$
Assume there exist maximizers $(p_n, \phi_n)$ of $L(p,\phi,Q_n)$, and a unique maximizer $(p^*,\phi^*)$ of $L(p,\phi,Q)$, where $p_n, p^*\in[0,1], \phi_n, \phi^*\in \tilde{\Phi}^d$. Let $f_n=exp(\phi_n)$, $f^*=exp(\phi^*)$, then
\begin{eqnarray}
  \stackrel[n\rightarrow\infty]{}{\textup{lim}}p_n &=& p^*,  \nonumber\\
  \stackrel[n\rightarrow\infty,\  x\rightarrow y]{}{\textup{lim}}f_n(x) &=& f^*(y), \ \ \forall y\in R^d\setminus \partial\{f^*>0\},  \nonumber\\
  \stackrel[n\rightarrow \infty, \ x\rightarrow y]{}{\textup{limsup}}f_n(x) &\leq& f^*(y), \ \  \forall y\in \partial\{f^*>0\},  \nonumber\\
  \stackrel[n\rightarrow\infty]{}{\textup{lim}}\int|f_n(x)-f^*(x)|dx &=& 0.  \nonumber
\end{eqnarray}
\end{thm}
Practically, $Q_n$ will be the empirical distribution function which automatically satisfies the above assumption.

\section{Simulation}\label{simulation}

In this section, we investigate the finite sample performance of our algorithm and compare it to the estimator proposed by \cite{patra2016estimation} ($\hat{\alpha}_0^{0.1k_n}$ from their paper), the Symmetrization estimator by \cite{bordes2006semiparametric}, the EM-type estimator, Maximizing-$\pi$ type estimator by \cite{song2010estimating}, and the Minimum profile Hellinger distance estimator by \cite{xiang2014minimum}.

In order to test our method under different settings, we simulate $K=200$ samples of $n$ i.i.d. random variables with the common distribution given by the following six models:
\begin{itemize}
\item Model 1: $g(x)=(1-p)*N(\mu=0,\sigma=2)+p*N(\mu=3,\sigma=1)$,
\item Model 2: $g(x)=(1-p)*\text{unif}(0,1)+p*\text{beta}(\alpha=1,\beta=5)$,
\item Model 3: $g(x)=(1-p)*\text{exp}(\lambda=1)+p*(\text{exp}(\lambda=1)+2)$,
\item Model 4: $g(x)=(1-p)*N(0,1)+p*(\chi^2(3)+2)$,
\item Model 5: $g(x)=(1-p)*N(0,1)+p*(\text{exp}(\lambda=0.5)+3)$,
\item Model 6: $g(x)=(1-p)*N(0,1)+p*(t(d.f.=5)+3)$.
\end{itemize}

For each sample we estimate $p$, the mean ($\mu$) of the unknown component $f$ and the classification error. The detailed calculation is explained bellow.

For our algorithm and the algorithm by \cite{xiang2014minimum}, final estimators $\hat{p}$ and $\hat{f}$ are always produced, thus the estimated probability $\hat{w}_i$ that the $i$-th observation is from the known component $f_0(x)$, given $X_i=x_i$, can be calculated by
$$\hat{w_i}=\frac{(1-\hat{p})f_0(x_i)}{(1-\hat{p})f_0(x_i)+\hat{p}\hat{f}(x_i)}.$$
For other methods, $\hat{f}$ may not always be given directly. Suggested by \cite{song2010estimating}, we estimate $\hat{w_i}$ by the following,
$$\hat{w_i}=\frac{2(1-\hat{p})f_0(x_i)}{(1-\hat{p})f_0(x_i)+\hat{h}(x_i)},$$
where $\hat{h}$ is the kernel density estimator of $g$ with Gaussian kernel and Silverman's ``rule of thumb'' bandwidth \citep{silverman1986density}. Note that the algorithm proposed by \cite{patra2016estimation} actually can estimate $\hat{f}$ when $f$ is non-increasing. But we find the algorithm works best when $f_0$ and $f$ has the same support and often produces unreliable estimates when two supports differ from each other. Thus, we do not use $\hat{f}$ to estimate $\hat{w_i}$ for \cite{patra2016estimation}'s algorithm even when the true $f$ does decrease on its support, instead, we follow \cite{song2010estimating}'s recommendation to get $\hat{w_i}$.

The algorithms by \cite{xiang2014minimum} and \cite{bordes2006semiparametric} give a final mean estimator $\hat{\mu}$ directly. For other methods, after we get $\hat{w_i}$, we estimate $\mu$ by the following weighted sum,
$$\hat{\mu}=\frac{\sum_{i=1}^n(1-\hat{w_i})X_i}{\sum_{i=1}^n (1-\hat{w_i})}.$$
Last, we report the classification error (Cla\_error) based on $\hat{w_i}$ as the mean squared error between $\hat{w_i}$ and the true $w_i$, i.e.,
 $$\text{Cla\_error}=\frac{1}{n}\sum_{i=1}^n(\hat{w_i}-w_i)^2,$$
 where $w_i=1$ if $x_i$ is from the known component $f_0(x)$ and $0$ if $x_i$ is from the unknown component $f(x)$.

For model 1, Table 1 reports the bias and MSE of the estimates of $p$, the bias  and MSE of the estimates of $\mu$, and the mean of the classification error for different methods over $K=200$ repetitions when $p=0.2$, $p=0.5$, and $p=0.8$, with sample size $n=1000$.  Similar reports of other models can be found in Tables \ref{model1} | \ref{model6}. Simulation results for sample sizes $n=250$ and $n=500$ are reported in the Appendix(Section \ref{appendix}). We report the results of \cite{bordes2006semiparametric}'s algorithm only for model 1, model 2 and model 6,  since this method fails to estimate $p$ for other models, in which the real $f(x)$'s are not symmetric on their supports.   \\

\begin{table}[htbp]

\caption{Bias(MSE) of estimates of  $p/\mu$ and mean of the classification error for model 1 when $n=1000$.}
	\label{model1}
\centering
\tabcolsep=0.08cm
\renewcommand*{\arraystretch}{1.3}
\begin{tabular}{c|cccccc}
  \hline
      &  EM\_logconcave &  Patra & Bordes &  Song EM &   Song max $\pi$  &   Xiang \\
  \hline
    \multicolumn{1}{c}{$p=0.2$} &   &   &   &  & & \\
  \hline
    \multicolumn{1}{c|}{$p$} & 0.002(0.0004) & 0.009(0.0007) & 0.001(0.0009) & 0.08(0.0066) & 0.087(0.0122) & 0.006(0.0006)\\
   \multicolumn{1}{c|}{$\mu$}  & 0.063(0.0180) & -0.152(0.0650) & -0.021(0.0426) & 0.116(0.0446) & -0.675(0.5680)& 0.116(0.0396) \\
   \multicolumn{1}{c|}{Cla\_error}  & 0.0960 & 0.1056 & 0.1052 & 0.1102 & 0.1052 & 0.0973 \\
  \hline
  \multicolumn{1}{c}{$p=0.5$} &   &   &   &  & & \\
  \hline
  \multicolumn{1}{c|}{$p$} & -0.002(0.0004)  & -0.025(0.0011)  & 0.001(0.0006)  & -0.132(0.0177) & 0.106(0.0149) & 0.007(0.0006)\\
   \multicolumn{1}{c|}{$\mu$}  & 0.018(0.0042)  & 0.051(0.0073)  & 0.000(0.0046) & 0.185(0.0375) &  -0.322(0.1392) & 0.013(0.0056)\\
   \multicolumn{1}{c|}{Cla\_error}  & 0.1094  & 0.1219  & 0.1198  & 0.1352 & 0.1206 & 0.1104 \\
   \hline
  \multicolumn{1}{c}{$p=0.8$} &   &   &   &  & & \\
  \hline
  \multicolumn{1}{c|}{$p$} & 0.001(0.0002)  & -0.252(0.0020)  & 0.001(0.0003)  & -0.107(0.0118) & 0.063(0.0047) & 0.009(0.0003) \\
   \multicolumn{1}{c|}{$\mu$}  & 0.005(0.0013)  & 0.066(0.0057)  & 0.000(0.0016) & 0.118(0.0153) &  -0.128(0.0220) & -0.002(0.0021)\\
   \multicolumn{1}{c|}{Cla\_error}  & 0.0645  & 0.0739  & 0.0694  & 0.0834 & 0.0721 & 0.0664 \\
  \hline
\end{tabular}
\end{table}

\begin{table}[htbp]
\caption{Bias(MSE) of estimates of  $p/\mu$ and mean of the classification error for model 2 when $n=1000$.}
\centering
\tabcolsep=0.08cm
\renewcommand*{\arraystretch}{1.3}
	\label{model2}
\begin{tabular}{c|cccccc}
  \hline
      &  EM\_logconcave &  Patra & Bordes &  Song EM &   Song max $\pi$  &   Xiang \\
   \hline
  \multicolumn{1}{c}{$p=0.2$} &   &   &   &  & & \\
  \hline
  \multicolumn{1}{c|}{$p$} &  -0.008(0.0014) & -0.023(0.0015)  & -0.015(0.0012)  & -0.15(0.0228) & 0.382(0.1496) & 0.017(0.0019) \\
   \multicolumn{1}{c|}{$\mu$}  & -0.018(0.0015)  & 0.027(0.0016)  & -0.029(0.0017) & -0.014(0.0007) & 0.199(0.0401) & -0.007(0.0010) \\
   \multicolumn{1}{c|}{Cla\_error}  &  0.1270 & 0.1520  & 0.1511  & 0.1676 & 0.1847 & 0.1339 \\
   \hline
  \multicolumn{1}{c}{$p=0.5$} &   & &     & &   &  \\
  \hline
  \multicolumn{1}{c|}{$p$} & 0.001(0.0007)  & -0.046(0.0030)  & -0.040(0.0024)   & -0.248(0.0811)  & 0.228(0.0548) & -0.047(0.0035)\\
   \multicolumn{1}{c|}{$\mu$}  & -0.003(0.0001)  &  -0.011(0.0002)  & -0.032(0.0011)  & -0.038(0.0015) & 0.077(0.0064) & -0.022(0.0012)\\
   \multicolumn{1}{c|}{Cla\_error}  & 0.1609  & 0.1990  & 0.1974   & 0.2638 & 0.1887 & 0.1753 \\
   \hline
  \multicolumn{1}{c}{$p=0.8$} &   &   &   &  & & \\
  \hline
  \multicolumn{1}{c|}{$p$} &  -0.001(0.0004) &  -0.074(0.0059)  & -0.070(0.0055) & -0.311(0.0974) & 0.099(0.0105) & -0.060(0.0043) \\
   \multicolumn{1}{c|}{$\mu$}  & -0.001(0.00004)  & -0.019(0.0004) & -0.033(0.0011)  & -0.040(0.0016) & 0.025(0.0007) & -0.030(0.0014) \\
   \multicolumn{1}{c|}{Cla\_error}  & 0.1000  & 0.1264 & 0.1261 & 0.2103  & 0.1129 & 0.1142  \\
  \hline
\end{tabular}
\end{table}

\begin{table}[htbp]
\caption{Bias(MSE) of estimates of  $p/\mu$ and mean of the classification error for model 3 when $n=1000$.}
\centering
\tabcolsep=0.08cm
\renewcommand*{\arraystretch}{1.3}
\label{model3}
\begin{tabular}{c|cccccc}
  \hline
      &  EM\_logconcave &  Patra & Bordes &  Song EM &   Song max $\pi$  &   Xiang \\
   \hline
  \multicolumn{1}{c}{$p=0.2$} &   &   &   &  & & \\
  \hline
  \multicolumn{1}{c|}{$p$} & 0.001(0.0002)  & -0.001(0.0006)  & NA  & -0.060(0.0038) & 0.410(0.1698) & 0.024(0.0011) \\
   \multicolumn{1}{c|}{$\mu$}  & 0.006(0.0082)  & -0.039(0.0152)  & NA & 0.048(0.0149) & -1.140(1.3094) & -0.105(0.0184) \\
   \multicolumn{1}{c|}{Cla\_error}  & 0.0709  & 0.0851  & NA  & 0.0879 & 0.1568 & 0.0790 \\
   \hline
  \multicolumn{1}{c}{$p=0.5$} &   &   &   &  & & \\
  \hline
  \multicolumn{1}{c|}{$p$} & 0.000(0.0003)  & -0.013(0.0006)  & NA  & -0.073(0.0057) & 0.259(0.0681) & 0.042(0.0028) \\
   \multicolumn{1}{c|}{$\mu$}  &  0.003(0.0021) & -0.011(0.0030) & NA & 0.018(0.0030) & -0.502(0.2578) & -0.091(0.0157) \\
   \multicolumn{1}{c|}{Cla\_error}  & 0.0595  & 0.0767  & NA & 0.0790 & 0.1166 & 0.0732 \\
   \hline
  \multicolumn{1}{c}{$p=0.8$} &   &   &   &  & & \\
  \hline
  \multicolumn{1}{c|}{$p$} & 0.001(0.0002)  & -0.228(0.0010)  &  NA  & -0.231(0.0012) & 0.104(0.0112) & 0.071(0.0060) \\
   \multicolumn{1}{c|}{$\mu$}  &  -0.001(0.0013) & -0.002(0.0014)  & NA & -0.002(0.0014) & -0.159(0.0283) & -0.104(0.0224) \\
   \multicolumn{1}{c|}{Cla\_error}  &  0.0260 & 0.0325  & NA  & 0.0322 & 0.0526 & 0.0617 \\
 \hline
\end{tabular}
\end{table}

\begin{table}[htbp]
\caption{Bias(MSE) of estimates of  $p/\mu$ and mean of the classification error for model 4 when $n=1000$.}
\centering
\tabcolsep=0.08cm
\renewcommand*{\arraystretch}{1.3}
	\label{model4}
\begin{tabular}{c|cccccc}
  \hline
      &  EM\_logconcave &  Patra & Bordes &  Song EM &   Song max $\pi$  &   Xiang \\
   \hline
  \multicolumn{1}{c}{$p=0.2$} &   &   &   &  & & \\
  \hline
  \multicolumn{1}{c|}{$p$} & -0.000(0.0002)  & 0.005(0.0005)  & NA  & 0.006(0.0003) & 0.106(0.0160) & 0.056(0.0041) \\
   \multicolumn{1}{c|}{$\mu$}  & -0.023(0.0321) & -0.286(0.1299)  & NA & -0.304(0.1353) & -1.066(1.4174) & -0.738(1.0279)\\
   \multicolumn{1}{c|}{Cla\_error}  & 0.0112  & 0.0139  & NA  & 0.0137 & 0.0205 & 0.0215 \\
   \hline
  \multicolumn{1}{c}{$p=0.5$} &   &   &   &  & & \\
  \hline
  \multicolumn{1}{c|}{$p$} & 0.000(0.0002)  & -0.009(0.0004)  & NA  & 0.014(0.0005) & 0.067(0.0057) & 0.049(0.0030) \\
   \multicolumn{1}{c|}{$\mu$}  & -0.005(0.0074)  & -0.148(0.0332)  & NA & -0.185(0.0459) & -0.333(0.1439) & -0.676(0.6616) \\
   \multicolumn{1}{c|}{Cla\_error}  & 0.0110  & 0.0157  & NA   & 0.0163 & 0.0160 & 0.0207 \\
   \hline
  \multicolumn{1}{c}{$p=0.8$} &   &   &   &  & & \\
  \hline
  \multicolumn{1}{c|}{$p$} & -0.001(0.0001)  & -0.023(0.0007)  & NA  & 0.006(0.0002) & 0.038(0.0019) & 0.066(0.0048) \\
   \multicolumn{1}{c|}{$\mu$}  &  -0.002(0.0052) & -0.025(0.0077)  & NA & -0.054(0.0098) & -0.147(0.0352) & -0.718(0.5396) \\
   \multicolumn{1}{c|}{Cla\_error}  & 0.0045  & 0.0078  & NA  & 0.0089 & 0.0108 & 0.0319 \\
   \hline
\end{tabular}
\end{table}

\begin{table}[htbp]
\caption{Bias(MSE) of estimates of  $p/\mu$ and mean of the classification error for model 5 when $n=1000$.}
\centering
\tabcolsep=0.08cm
\renewcommand*{\arraystretch}{1.3}
\label{model5}
\begin{tabular}{c|cccccc}
  \hline
      &  EM\_logconcave &  Patra & Bordes &  Song EM &   Song max $\pi$  &   Xiang \\
   \hline
  \multicolumn{1}{c}{$p=0.2$} &   &   &   &  & & \\
  \hline
  \multicolumn{1}{c|}{$p$} & 0.001(0.0002)  & 0.006(0.0006)  & NA  & 0.018(0.0005) & 0.110(0.0154) & 0.037(0.0018) \\
   \multicolumn{1}{c|}{$\mu$}  & 0.004(0.0193)  & -0.399(0.2021)  & NA & -0.427(0.2126) & -1.129(1.4965) & -0.923(0.9073)\\
   \multicolumn{1}{c|}{Cla\_error}  & 0.0012  & 0.0044  & NA  & 0.0045 & 0.0105 & 0.0092 \\
   \hline
  \multicolumn{1}{c}{$p=0.5$} &   &   &   &  & & \\
  \hline
  \multicolumn{1}{c|}{$p$} & 0.000(0.0003)  & -0.009(0.0005)  & NA  & 0.023(0.0008) & 0.131(0.0208) & 0.061(0.0044) \\
   \multicolumn{1}{c|}{$\mu$}  & -0.001(0.0079)  & -0.173(0.0384)  & NA & -0.213(0.0538) & -0.681(0.5620) & -0.650(0.4645)    \\
   \multicolumn{1}{c|}{Cla\_error}  & 0.0007  & 0.0079  & NA  & 0.0090 & 0.0202 & 0.0189 \\
   \hline
  \multicolumn{1}{c}{$p=0.8$} &   &   &   &  & & \\
  \hline
  \multicolumn{1}{c|}{$p$} & 0.001(0.0001)  & -0.223(0.0007)  &  NA & 0.009(0.0002)  & 0.079(0.0068) & 0.081(0.0071) \\
   \multicolumn{1}{c|}{$\mu$}  & 0.000(0.0047)  & -0.027(0.0061)  & NA & -0.051(0.0077) & -0.313(0.1123) & -0.473(0.2482) \\
   \multicolumn{1}{c|}{Cla\_error}  & 0.0003  & 0.0022  & NA  & 0.0030 & 0.0190 & 0.0426 \\
  \hline
\end{tabular}
\end{table}

\begin{table}[htbp]
\caption{\label{model6} Bias(MSE) of estimates of  $p/\mu$ and mean of the classification error for model 6 when $n=1000$.}
\centering
\tabcolsep=0.08cm
\renewcommand*{\arraystretch}{1.3}
\begin{tabular}{c|cccccc}
  \hline
      &  EM\_logconcave &  Patra & Bordes &  Song EM &   Song max $\pi$  &   Xiang \\
   \hline
  \multicolumn{1}{c}{$p=0.2$} &   &   &   &  & & \\
  \hline
  \multicolumn{1}{c|}{$p$} & -0.010(0.0003)  & -0.007(0.0006)  & 0.083(0.0012)  & -0.022(0.0006) & 0.077(0.0080) & 0.001(0.0003) \\
   \multicolumn{1}{c|}{$\mu$}  & 0.171(0.0455)  & 0.029(0.0262)  & -0.075(0.0843) & 0.083(0.0257) & -0.414(0.2369) & 0.001(0.0177)\\
   \multicolumn{1}{c|}{Cla\_error}  & 0.0440  & 0.0450  & 0.0455  & 0.0457 & 0.0468 & 0.0435 \\
   \hline
  \multicolumn{1}{c}{$p=0.5$} &   &   &   &  & & \\
  \hline
  \multicolumn{1}{c|}{$p$} & 0.001(0.0003)  & -0.031(0.0015)  & -0.002(0.0006)  & -0.053(0.0031) & 0.038(0.0028) & -0.002(0.0631) \\
   \multicolumn{1}{c|}{$\mu$}  & -0.010(0.0115)  & 0.148(0.0264)  & -0.003(0.0066) & 0.182(0.0375) & -0.020(0.0185) & 0.018(0.0066) \\
   \multicolumn{1}{c|}{Cla\_error}  & 0.0094  & 0.0672  & 0.0658  & 0.0680 & 0.0656 & 0.0631 \\
   \hline
  \multicolumn{1}{c}{$p=0.8$} &   &   &   &  & & \\
  \hline
  \multicolumn{1}{c|}{$p$} & -0.001(0.0001)  & -0.059(0.0037)  & -0.002(0.0004)  & -0.063(0.0043) & 0.008(0.0006) & 0.001(0.0034)\\
   \multicolumn{1}{c|}{$\mu$}  & -0.004(0.0072)  & 0.169(0.0307)  & -0.001(0.0024) & 0.174(0.0321) & 0.055(0.0073) & -0.003(0.0034) \\
   \multicolumn{1}{c|}{Cla\_error}  & 0.0046  & 0.0637  & 0.0570  & 0.0643 & 0.0567 & 0.0545\\
  \hline
\end{tabular}
\end{table}

All the simulation results strongly suggest that our method is very competitive and often outperforms all other methods. 
Moreover, our method is even more favorable when the sample size $n$ gets larger.

\begin{figure}[H]
  \centering
  \includegraphics[width=7cm,height=7cm]{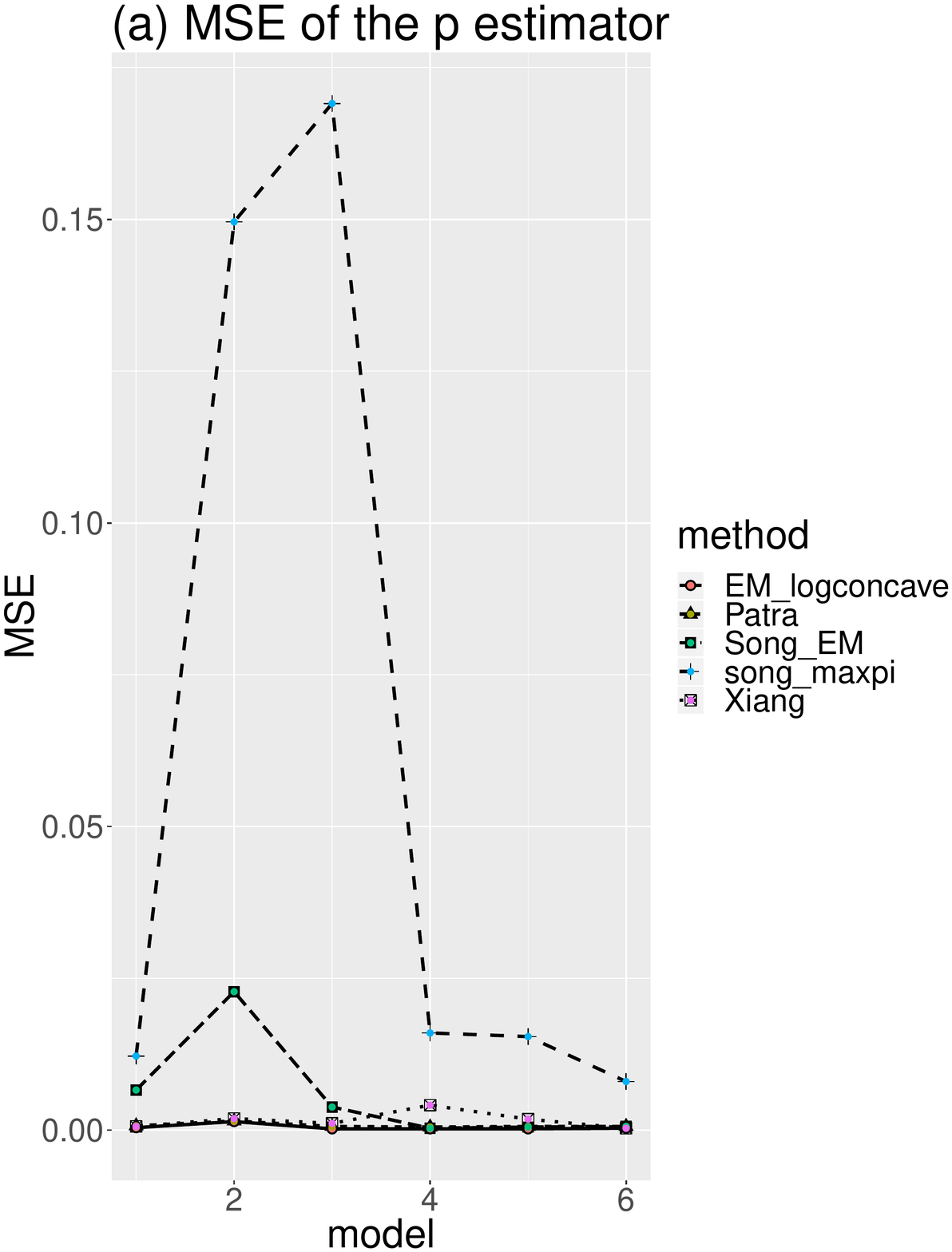} \includegraphics[width=7cm,height=7cm]{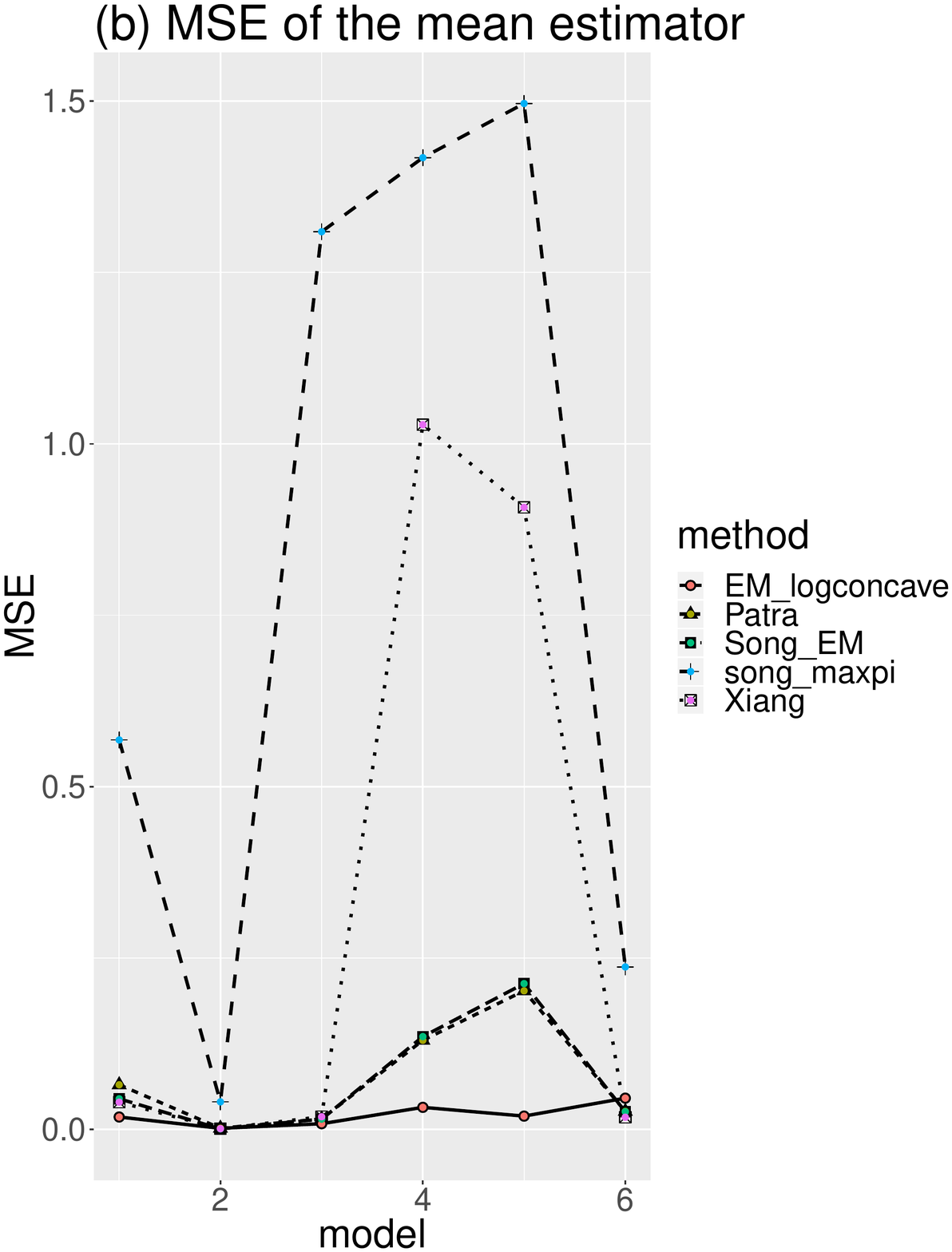}\\
  \caption{(a): MSE of the estimates of $p$ when $p=0.2, \ n=1000$; (b): MSE of the estimates of $\mu$ when $p=0.2, \ n=1000$. }
  \label{mse}
\end{figure}

To better display our simulation results, we also plot the MSE of point estimates of $p$ and $\mu$ vs. different models for all the methods we mentioned above when $p=0.2$ and $n=1000$,  except for the method by \cite{bordes2006semiparametric} as their method fails to estimate $p$ and $\mu$ for half of the models we discussed here. Figure \ref{mse} shows that the curve representing our method always lies at the bottom which demonstrates the effectiveness of our algorithm, while the Maximizing-$\pi$ type estimator by \cite{song2010estimating} gives the worst results in terms of MSE.

\hfill\\

\section{Real Data Application}\label{realdata}
\subsection{Prostate Data}

In this section we consider the prostate data consisting of genetic expression levels related to prostate cancer patients of \cite{efron2012large}. The data set is a $6033\times 102$ matrix, with entries $x_{ij}=$ expression level for gene $i$ on patient $j$, $i=1,\cdots, n$, $j=1,\cdots, m$, here $n=6033$, $m=102$. Among the $m=102$ patients, $m_1=50$ of them are normal control subjects (corresponding to $j=1,\cdots, m_1$) and $m_2=52$ of them are prostate cancer patients (corresponding to $j=m_1+1,\cdots,m_2$). The goal of the study is to discover the potential genes that are differentially expressed between normal control and prostate cancer patients.

Two-sample $t$-test is performed to test the significance of each gene $i$ by,
 $$t_i=(\bar{x}_i(1)-\bar{x}_i(2))/{s_i},$$ where $\bar{x}_i(1)=(\stackrel[j=1]{m_1}{\sum}x_{ij})/m_1$,   $\bar{x}_i(2)=(\stackrel[j=m_1+1]{m_2}{\sum}x_{ij})/m_2$,
$s_i^2=(1/m_1+1/m_2)\left\{\stackrel[j=1]{m_1}{\sum}(x_{ij}-\bar{x}_i(1))^2+\right.$
$\left.\stackrel[j=m_1+1]{m_2}{\sum}(x_{ij}-\bar{x}_i(2))^2\right\}/(m-2)$.
These two-sided $t$-tests produce $n=6033$ $p$-values, and the distribution of these $p$-values under the null hypothesis (i.e., gene $i$ is not differentially expressed) has a uniform density,  while under the alternative hypothesis (i.e., gene $i$ is differentially expressed) has a non-increasing density.

The estimation of $p$ is reported in Table \ref{tab:prostate}. We can see that the estimate by \cite{bordes2006semiparametric} and the Maximizing-$\pi$ type estimate by \cite{song2010estimating} give a relatively big estimate. The estimate procedure by \cite{bordes2006semiparametric} assumes the density function under the alternative hypothesis to be symmetric, while in our example this density is non-increasing, which implies the violation of their symmetric assumption. It is known that the Maximizing-$\pi$ type estimator by \cite{song2010estimating} tends to overestimate the $p$ value, which can also been seen in Table \ref{tab:prostate}. We also want to point out that several approaches have been proposed by \cite{efron2012large} to estimate $p$ as well, the estimator based on central matching method gives $\hat{p}=0.020$ (please see \cite{efron2012large} and \cite{efron2007size} for detailed description of those estimators), and Table \ref{tab:prostate} shows that our estimator gives a closest value to Efron's result.\\

\begin{table}[htbp]
	\caption{Estimates of $p$ for the prostate cancer data.}
	\centering
	\tabcolsep=0.08cm
	\renewcommand*{\arraystretch}{1.3}
	\label{tab:prostate}
	\begin{tabular}{cccccc}
	 \hline
	EM\_logconcave & \text{Patra} & \text{Bordes} & \text{Song EM}   & \text{Song max}\  $\pi$    & \text{Xiang}\\
	\hline
	0.0173 & 0.0817 & 0.1975 & 0.0076 & 0.6132 & 0.1915\\
	\hline	
\end{tabular}
\end{table}
	

\vspace{0.5cm}

\begin{figure}[H]
  \centering
  \includegraphics[width=6cm,height=6cm]{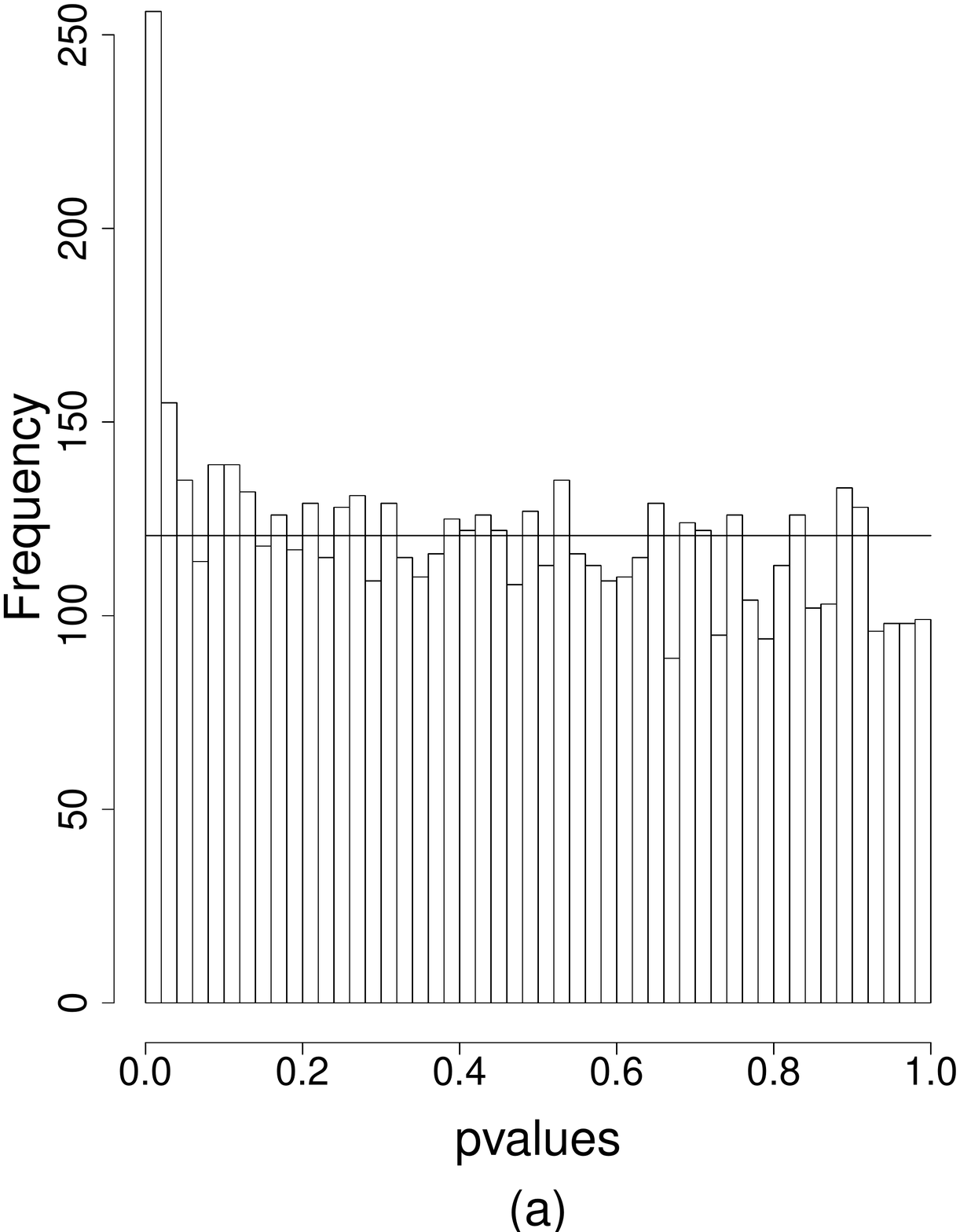}\\
  \includegraphics[width=6cm,height=6cm]{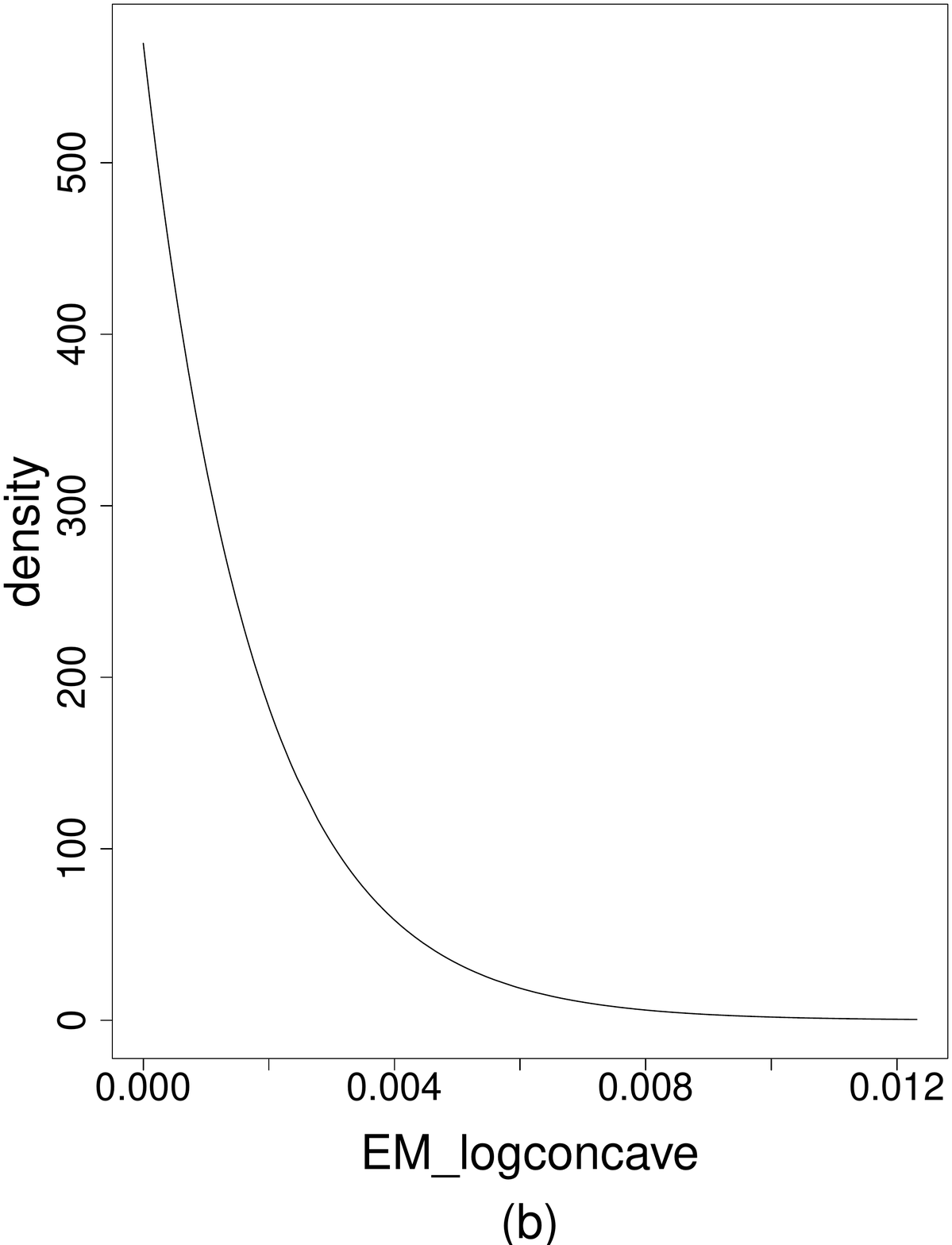}
  \includegraphics[width=6cm,height=6cm]{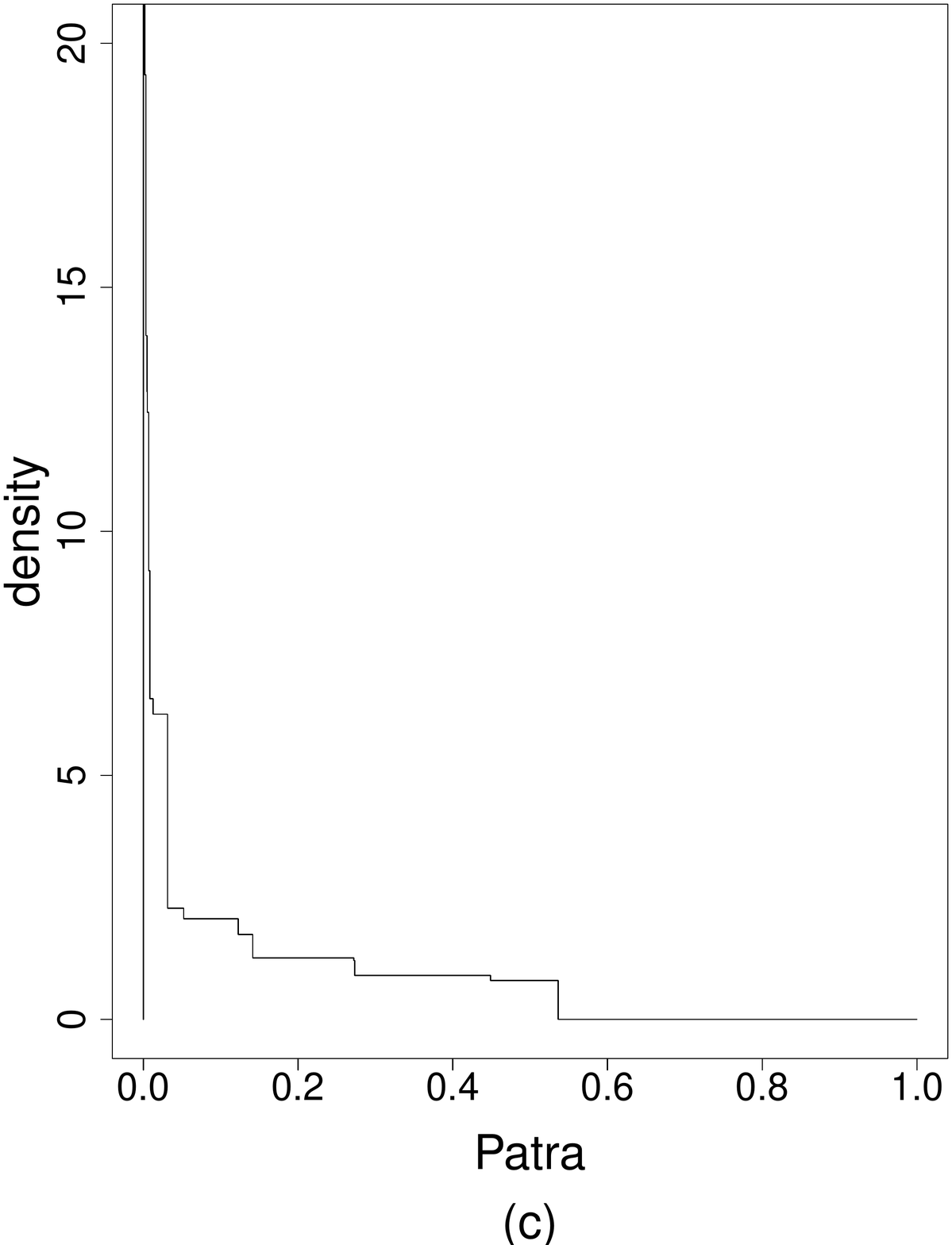}\\
  \caption{Plots for the prostate data: (a) Histogram of the $p$-values. The horizontal line represents the Uniform(0,1) distribution. (b) Plot of the estimated density $\hat{f}$ by our maximum likelihood estimation via EM algorithm. (c) Plot of the estimated density $\hat{f}$ by the method of \cite{patra2016estimation}. }\label{prostate}
\end{figure}

Figure \ref{prostate} shows that our estimate of the density $\hat{f}$ under the alternative tends to have a much smaller support comparing to the one given by \cite{patra2016estimation}. Again, as we noted before, the method by \cite{patra2016estimation}  assumes that $f$ is decreasing on the whole support of $f_0$. While in reality, smaller $p$-values tends to indicate the alternative hypothesis, hence it actually makes sense that the support of $f$ for this prostate data may be much smaller than $(0,1)$. The estimate produced by \cite{bordes2006semiparametric} is not very reliable, since the density $f$ is not symmetric. For this example, if we apply \cite{bordes2006semiparametric}'s method to the original $t$ statistics directly, the estimate is $\hat{p}=0.0072$.


\hfill\\

\subsection{Carina Data}

Carina is one of the seven dwarf spheroidal (dSph) satellite galaxies of Milkey Way. Here we consider the data consisting of radial velocities (RV) of $n=1266$ stars from Carina galaxy. The data is obtained by Magellan and MMT telescopes (\cite{walker2007velocity}). The stars of Milkey Way contribute contamination to this data set. We assume the distribution $f_0$ of RV from stars of Milkey Way is known from the Besancon Milky Way model (\cite{robin2003synthetic}). Now we would like to analyze this data set to better understand the mixture distribution of the RV of stars in Carina galaxy.

The estimation of $p$ is reported in Table \ref{tab:carina}. Again we see that the estimation by \cite{song2010estimating}'s Maximizing-$\pi$ type estimator gives a relatively big estimate. Other estimates are relatively close.\\

\begin{table}[htbp]
	\caption{Estimates of $p$ for the Carina data.}
	\centering
	\tabcolsep=0.08cm
	\renewcommand*{\arraystretch}{1.3}
	\label{tab:carina}
	\begin{tabular}{cccccc}
	 \hline
	EM\_logconcave & \text{Patra} & \text{Bordes} & \text{Song EM}   & \text{Song max}\  $\pi$    & \text{Xiang}\\
	\hline
	  0.354 & 0.364 & 0.363 & 0.370 & 0.687 & 0.385\\
	\hline	
\end{tabular}
\end{table}



\begin{figure}
  \centering
  \includegraphics[width=8cm,height=8cm]{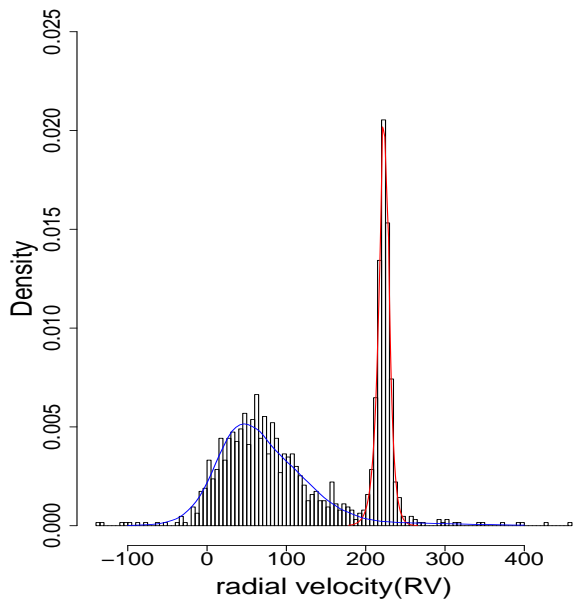}\\
  \caption{Histogram of RV data overlaid with the estimated two components from our EM log-concave algorithm}\label{carina_logconcave}
\end{figure}

Next in Figure \ref{carina_logconcave} we plot the histogram of the RV data overlaid with our estimated two components of the mixture density, and we can see that our estimation approximates the real data fairly well. The component corresponding to the stars of Carina looks very symmetric, and in fact astronomers usually assume the distribution to be Gaussian, which causing the density estimation proposed by \cite{patra2016estimation} does not work here.

\hfill\\

\section{Discussion}\label{discussion}

In this paper we study the two-component mixture model with one component completely known.  A nonparametric maximum likelihood estimator is developed via EM algorithm and log-concave approximation. Unlike most existing estimation procedures, our new method finds the mixing proportion and the distribution of the unknown component simultaneously without any selection of a tuning parameter and the proposed EM algorithm satisfies the non-decreasing property of the traditional EM algorithm. Simulation results show that our method is more favorable than many other competing estimation methods. 

We are able to prove the existence and consistency of our maximum likelihood estimator for a general distribution $Q$. But we do require some extra conditions on $f_0$ and $f$. A possible future research direction is trying to ease these assumptions and make it more general. In addition, it would be also our interest to apply our method to a more general model where the component $f_0$ also contains some unknown parameter. \\

\section{Appendix}\label{appendix}

\subsection{Theoretical Proof}

\begin{proof}[\bf Proof of Proposition \ref{identifiability1}]
According to \cite{patra2016estimation}, if we let $G,\ F_0$, and  $F$ be the cumulative distribution functions of $g, \ f_0$ and $f$ respectively, define $p_0=\text{inf}\{\gamma\in(0,1]: [G-(1-\gamma)F_0]/\gamma \ \text{is a CDF}\}$, then $$p_0=p\{1-\text{essinf}\frac{f}{f_0}\},$$
where $\text{essinf}(h)=\text{sup}\{t\in R: \mathfrak{m}\{x: h(x)<t\}=0\}$, and here $\mathfrak{m}$ represents the Lebesgue measure. Now if $\text{essinf}\frac{f}{f_0}>0$, there must exist some $t>0$, such that,  $\mathfrak{m}\{x: \frac{f(x)}{f_0(x)}<t\}=0$, i.e., $\frac{f(x)}{f_0(x)}\geq t$ almost everywhere, which contradicts to the fact that $\lim_{x\rightarrow a^+}\frac{f(x)}{f_0(x)}=0  \ \text{or} \
\lim_{x\rightarrow a^-}\frac{f(x)}{f_0(x)}=0 $. Hence we can conclude that $\text{essinf}\frac{f}{f_0}=0$, and consequently $p_0=p$, which means if we can write $g(x)=(1-p)f_0(x)+pf(x)$, this $p$ is fixed and equals $p_0$. Consequently $f(x)=(g(x)-(1-p)f_0(x))/p$ is fixed as well, and our model (\ref{model}) is identifiable.
\end{proof}

\begin{proof}[\bf Proof of Proposition \ref{identifiability2}]
 Since $\displaystyle f(x)=e^{\phi(x)}$ is a log-concave density, there exist constants $a$ and $b>0$, such that $\phi(x)\leq a-b|x|$ (see \cite{cule2010theoretical}), which implies
$$\phi(x)-\text{log}f_0(x)\leq a-b|x|-\text{log}f_0(x).$$
Now if $\text{log}f_0(x)=O(x^k)$, for some $0<k<1$, apparently,
$$-b|x|-\text{log}f_0(x)=
|x|^k(-b|x|^{1-k}-\text{log}f_0(x)/|x|^k)\rightarrow -\infty,\ \text{as} \ x\rightarrow +\infty \ \text{or}\  x\rightarrow -\infty.
$$
 Hence $\phi(x)-\text{log}f_0(x)\rightarrow -\infty$ as $x\rightarrow +\infty$ or $x\rightarrow -\infty$, which shows $\lim_{x\rightarrow +\infty}\frac{f(x)}{f_0(x)}=0$. Thus, model (\ref{model}) is identifiable from Proposition \ref{patra}.
\end{proof}

\begin{proof}[\bf Proof of Theorem \ref{existence}]
 Suppose $\int ||x||^kQ(dx)<\infty$, $\text{interior}(\text{csupp}(Q))\neq\emptyset$, $\int e^{\phi(x)}dx=1$ and $f_0(x)\leq m(x)e^{\phi(x)}$. For any concave function $\phi$ satisfying the above conditions, there exist $(a_0, b_0)$, such that $\phi(x)\leq a_0-b_0||x||$, thus for any $p\neq 0$, $L(p,-b_0||x||-\text{log}(\int e^{-b_0||x||}dx),Q)\geq \text{log} \frac{p}{\int e^{-b_0||x||}dx}-b_0\int ||x||Q(dx)>-\infty$,  thus we have $L(Q)>-\infty$. When maximizing $L(p,\phi,Q)$ over all $\phi\in\tilde{\Phi}^d$, we may restrict our attention to functions $\phi$ such that $\text{dom}(\phi)=\{x\in R^d: \phi(x)>-\infty\}\subseteq \text{csupp}(Q)$. For if $\text{dom}(\phi)\nsubseteq \text{csupp}(Q)$, replacing $\phi(x)$ with $-\infty$ for all $x\notin \text{csupp}(Q)$, then the value of $L(p, \phi-\text{log}(\int e^{\phi(x)}dx), Q)$ would be greater or equal to the original $L(p, \phi, Q)$. Note that since $\text{csupp}(f_0)\subseteq \text{csupp}(Q)$, the new concave function $\phi'=\phi-\text{log}(\int e^{\phi(x)}dx)$ still satisfies the conditions above, i.e., $\int e^{\phi'(x)}dx=1$, $f_0(x)\leq m(x)e^{\phi'(x)}$ and $\text{dom}(\phi')=\{x\in R^d: \phi(x)>-\infty\}\subseteq \text{csupp}(Q)$. We denote $\Phi(Q)$ to be the family of all $\phi\in\Phi^d$ with these properties.

Now we show that $L(Q)<\infty$. Suppose that $\phi\in \Phi(Q)$ is such that $M=\text{max}_{x\in R^d}\phi(x)>0$. Let $D_t=\{\phi\geq t\}$, hence $D_t$ is closed and convex. For any $\alpha>0$, we have the following estimate,
\begin{eqnarray}
  L(p,\phi,Q)  &=& \int \text{log}((1-p)f_0+pe^\phi)dQ   \nonumber\\
    &\leq& \int \text{log}((1-p)m(x)+p)Q(dx)+\int\phi dQ   \nonumber\\
    &\leq& \int \text{log}(m(x)+1)Q(dx)-\alpha MQ(R^d \setminus D_{-\alpha M})+MQ(D_{-\alpha M})  \nonumber\\
    &=& \int \text{log}(m(x)+1)Q(dx)-(\alpha+1)M(\frac{\alpha}{\alpha+1}-Q(D_{-\alpha M})).\nonumber
\end{eqnarray}

Note that $\int \text{log}(m(x)+1)Q(dx)$ exists since $\int ||x||^k Q(dx)<\infty$. By Lemma 4.1 of \cite{dumbgen2011approximation}, for any fixed $\alpha$,
\begin{eqnarray}
  Leb(D_{-\alpha M}) &\leq& (1+\alpha)^dM^de^{-M}/\int_0^{(1+\alpha)M}t^de^{-t}dt  \nonumber\\
   &=& (1+\alpha)^dM^de^{-M}/(d!+o(1))\rightarrow 0, \  \text{as} \  M\rightarrow \infty. \nonumber
\end{eqnarray}

Lemma 2.1 of \cite{dumbgen2011approximation} says that for sufficiently large $\alpha$ and sufficiently small $\delta>0$, there exist some sufficiently small $\epsilon>0$, such that,
$$\text{sup}\{Q(C): C\subseteq R^d \ \text{closed and convex},\ \text{Leb}(C)\leq \delta\}<\frac{\alpha}{\alpha+1}-\epsilon,$$
which implies that $L(p,\phi,Q)\rightarrow-\infty$, as $M\rightarrow\infty$. Since for any $\phi\in\Phi(Q)$, we also have $L(p,\phi,Q)\leq \int \text{log}((1-p)m(x)+p)dQ+M$, $\Rightarrow L(Q)<\infty$ and there exist constants $M_0$ and $M_*$, such that\\ $  L(Q)=\stackrel[{\stackrel[ M_0\leq \text{max}(\phi(x))\leq M_*]{}{p\in[0,1], \ \phi\in\Phi(Q)}}]{}{\text{sup}}L(p,\phi,Q)$.

Now that we know $L(Q)$ is real, we are ready to prove the existence of a maximizer $(p_0,\phi_0)$ of $L(Q)$. Let $(p_n,\phi_n)$ be a sequence such that $p_n\in[0,1]$, $\phi_n\in\Phi(Q)$, $M_n=\text{max}(\phi_n(x))\in[M_0, M_*]$, and $-\infty<L(p_n, \phi_n,Q)\uparrow L(Q)$ as $n\rightarrow\infty$. Here we assume $\{p_n\}$ is a convergent sequence, say $p_n\rightarrow p_0\in[0,1]$, as $n\rightarrow\infty$. If $\{p_n\}$ is not convergent, since it is bounded, it must have a convergent subsequence $\{
p_{n_k}\}$, and the sequence $\{p_{n_k},\phi_{n_k}\}$ would satisfy all those properties above and we can just simply replace the original sequence with this subsequence.

Next, we show that,
\begin{equation}\label{infimumofphi}
\stackrel[n\geq 1]{}{\text{inf}}\phi_n(x_0)>-\infty, \ \forall x_0\in \text{interior}(\text{csupp}(Q)).
\end{equation}

For any $x_0\in \text{interior}(\text{csupp}(Q))$, if $\phi_n(x_0)<M_n$, then $x_0$ can not be an interior point of $\{\phi_n\geq \phi_n(x_0)\}$, hence,
\begin{eqnarray}
    L(p_n,\phi_n,Q)&=& \int \text{log}((1-p_n)f_0+p_ne^{\phi_n})dQ  \nonumber\\
    &\leq& \int \text{log}(m(x)+1)Q(dx)+\int \phi_n dQ  \nonumber\\
    &\leq& \int \text{log}(m(x)+1)Q(dx)+\phi_n(x_0)+(M_n-\phi_n(x_0))Q(\phi_n\geq\phi_n(x_0)) \nonumber\\
    &\leq& \int \text{log}(m(x)+1)Q(dx)+\phi_n(x_0)(1-h(Q,x_0))+\text{max}(M_n,0), \nonumber
\end{eqnarray}
where $h(Q,x)=\text{sup}\{Q(C): C\subseteq R^d \ \text{closed and convex}, \  x\notin \text{interior}(C)\}<1$ by Lemma 2.13 of \cite{dumbgen2011approximation}. And the above inequalities still hold even if $\phi_n(x_0)=M_n$. Thus we have,
\begin{eqnarray}
  \phi_n(x_0) &\geq& \frac{L(p_n,\phi_n,Q)-\int \text{log}(m(x)+1)Q(dx)-\text{max}(M_n,0)}{1-h(Q,x_0)} \nonumber\\
  \Rightarrow \stackrel[n\geq 1]{}{\text{inf}}\phi_n(x_0) &\geq& \frac{L(p_1,\phi_1,Q)-\int \text{log}(m(x)+1)Q(dx)-\text{max}(M_*,0)}{1-h(Q,x_0)}>-\infty, \nonumber
\end{eqnarray}
which establishes (\ref{infimumofphi}). Since $\phi_n\leq M_*$, together with (\ref{infimumofphi}), Lemma 3.3 of \cite{schuhmacher2011multivariate} implies that there exist constants $a$ and $b>0$ such that,
\begin{equation}\label{commonbound}
\phi_n(x)\leq a-b||x||, \ \forall n\geq 1, x\in R^d.
\end{equation}
Let $C=\{x\in R^d: \stackrel[n\rightarrow\infty]{}{\text{liminf}}\phi_n(x)>-\infty\}\supseteq \text{interior}(\text{csupp}(Q))$ and $\bar{\phi}(x)=a-b||x||$, using Lemma 4.2 of \cite{dumbgen2011approximation}, together with (\ref{infimumofphi}) and (\ref{commonbound}) we can conclude that there exist $\phi_0\in\Phi^d$ and a subsequence $\phi_{n_k}$ such that $C\subseteq \text{dom}(\phi_0)\subseteq \text{csupp}(Q)$ and,
\begin{eqnarray}
  \stackrel[k\rightarrow\infty]{}{\text{limsup}}\phi_{n_k}(x) &\leq& \phi_0(x)\leq a-b||x||, \  \forall x\in R^d,  \nonumber\\
  \stackrel[k\rightarrow\infty]{}{\text{lim}}\phi_{n_k}(x) &=& \phi_0(x)>-\infty, \ \forall x\in \text{interior}(\text{csupp}(Q)).  \nonumber
\end{eqnarray}

Since $\text{dom}(\phi_{n_k})\subseteq \text{csupp}(Q)$, we have $\phi_{n_k}$ converges to $\phi_0$ almost everywhere as the Lebesgue measure of the boundary of $\text{csupp}(Q)$ is zero, then we can conclude $\int e^{\phi_0(x)}dx=1$ by dominated convergence. Thus, $\phi_0\in \Phi(Q)$. Next, we apply Fatou's Lemma to the nonnegative functions $x\mapsto \int \text{log}(m(x)+1)Q(dx)+a-b||x||-\text{log}((1-p_{n_k})f_0+p_{n_k}e^{\phi_{n_k}})$, and we get,
$$\stackrel[k\rightarrow \infty]{}{\text{limsup}}L(p_{n_k},\phi_{n_k},Q)\leq L(p_0,\phi_0,Q).$$
Hence,
$$L(Q)\geq L(p_0,\phi_0,Q)\geq \stackrel[k\rightarrow \infty]{}{\text{limsup}}L(p_{n_k},\phi_{n_k},Q)=L(Q),$$
which shows $(p_0,\phi_0)$ is the maximizer that we are looking for.\\
\end{proof}

\begin{proof}[\bf Proof of Theorem \ref{consistency}]
Since $\stackrel[n\rightarrow \infty]{}{\text{lim}}D_k(Q_n, Q)\rightarrow 0$, hence
$$Q_n\rightarrow_w Q \ \text{and}   \int ||x||^k Q_n(dx)\rightarrow \int ||x||^kQ(dx), \ \text{as} \ n\rightarrow \infty.$$

Suppose $\stackrel[n\rightarrow \infty]{}{\text{limsup}}L(Q_n)=\lambda\in [-\infty, \infty]$, thus there exist a subsequence $\{Q_{n_k}\}$, such that $L(Q_{n_k})\rightarrow \lambda$. If we let $h(x)=-b_0||x||-\text{log}(\int e^{-b_0||x||}dx)$ as we did in the proof of Theorem \ref{existence}, then, $h\in\tilde{\Phi}^d$,  and for any $p>0$,
\begin{eqnarray}
\lambda &\geq & \stackrel[k\rightarrow\infty]{}{\text{limsup}}L(p,h,Q_{n_k})=\stackrel[k\rightarrow\infty]{}{\text{limsup}}\int \text{log}((1-p)f_0+pe^{h})dQ_{n_k} \nonumber\\
&\geq &
\text{log}p-b_0\int||x||Q(dx)-\text{log}(\int e^{-b_0||x||}dx)>-\infty.\nonumber
\end{eqnarray}

Note that in the above inequalities, we used the fact that $\stackrel[n\rightarrow\infty]{}{\text{lim}}\int||x||Q_n(dx)=\int||x||Q(dx)$ by Lemma 4.6 of \cite{schuhmacher2011multivariate}.

Let $M_n=\text{max}_{x\in R^d}\phi_n(x)$. Since $\stackrel[n\rightarrow\infty]{}{\text{lim}}\int \text{log}(m(x)+1)Q_n(dx)=\int \text{log}(m(x)+1)Q(dx)$ by Lemma 4.6 of \cite{schuhmacher2011multivariate}, similar to the proof of Theorem \ref{existence},  one can show that for $n$ sufficiently large, we have $L(p_n, \phi_n,Q_n)\rightarrow-\infty$, if $M_n\rightarrow\infty$ as $n\rightarrow\infty$, and $L(p_n,\phi_n, Q_n)\leq \int \text{log}(m(x)+1)Q(dx)+M_n$, provided that
\begin{equation}\label{limsup}
\stackrel[n\rightarrow \infty]{}{\text{limsup}}Q_n(C_n)<1, \  \text{for any} \  \{C_n: C_n\subseteq R^n \ \text{closed and convex}, \ \stackrel[n\rightarrow\infty]{}{lim}\text{Leb}(C_n)=0\}.
\end{equation}
Hence there exist some suitable constants $M_0$ and $M_*$, such that $M_0<M_{n_k}<M_*$ for $k$ sufficiently large and thus $\lambda<\infty$.

Here we explain how (\ref{limsup}) is derived. As in the proof of Lemma 2.1 of Schuhmacher, \cite{schuhmacher2011multivariate}, there exist a simplex $\tilde{\Delta}=\text{conv}(\tilde{x}_0, \cdots, \tilde{x}_d)$ with positive Lebesgue measure and open sets $U_0$, $U_1$, $\cdots$, $U_d$ with $Q(U_j)\geq\eta>0$, for $0\leq j\leq d$, here $\eta=\stackrel[0\leq j\leq d]{}{\text{min}}Q(U_j)>0$. For any convex and closed set $C$ with $C\cap U_j\neq \emptyset$ for all $j$, we have $\tilde{\Delta}\subseteq C$. By Theorem 4.4.4 of \cite{chung2001course}, $\stackrel[n\rightarrow\infty]{}{\text{liminf}}Q_n(U_j)\geq Q(U_j)\geq\eta$ for all $j$. Thus if $\stackrel[n\rightarrow\infty]{}{\text{lim}}\text{Leb}(C_n)=0$, then for any $n$ sufficiently large, $\text{Leb}(C_n)<\text{Leb}(\tilde{\Delta}),\Rightarrow \tilde{\Delta}\nsubseteq C_n,\Rightarrow$ there exist some $j$, such that $C_n\cap U_j=\emptyset,\Rightarrow Q_n(C_n)\leq 1-Q_n(U_j)\leq 1-\stackrel[1\leq j\leq d]{}{\text{min}}Q_n(U_j)$. Since,
\begin{eqnarray}
Q_n(U_j)&=&\stackrel[n\rightarrow\infty]{}{\text{liminf}} Q_n(U_j)+Q_n(U_j)-\stackrel[n\rightarrow\infty]{}{\text{liminf}}Q_n(U_j)\nonumber\\
&\geq& \eta+\stackrel[k\geq n]{}{\text{inf}}Q_k(U_j)-\stackrel[n\rightarrow \infty]{}{\text{liminf}}Q_n(U_j)=\eta+o(1),\nonumber
\end{eqnarray}
thus $\stackrel[1\leq j\leq d]{}{\text{min}}Q_n(U_j)\geq \eta+o(1)$, which shows that $Q_n(C_n)\leq 1-\eta+o(1)$, and hence (\ref{limsup}) is established.

Now that we know $M_{n_k}$ is bounded for $k$ sufficiently large, and $L(p_{n_k}, \phi_{n_k},Q_{n_k})\rightarrow\lambda\in R$ as $k\rightarrow\infty$, we may assume $\{p_{n_k}\}$, is a convergent sequence, say $p_{n_k}\rightarrow p_*\in[0,1]$, as $k\rightarrow\infty$. For if $\{p_{n_k}\}$ is not convergent, since it is bounded, it must have a convergent subsequence $\{p_{n_{k_l}}\}$, and the sequence $\{p_{n_{k_l}},\phi_{n_{k_l}}\}$ would satisfy all those properties above and we can just simply replace the original sequence with this subsequence.

Again, as in the proof of Theorem \ref{existence}, for any $x_0\in \text{interior}(\text{csupp}(Q))$, we have,
\begin{eqnarray}
  \phi_{n_k}(x_0) &\geq& \frac{L(p_{n_k},\phi_{n_k},Q_{n_k})- \int \text{log}(m(x)+1)Q(dx)-\text{max}(M_{n_k},0)}{1-h(Q_{n_k},x_0)}. \nonumber
\end{eqnarray}
As Lemma 2.13 of \cite{schuhmacher2011multivariate} states that $\stackrel[n\rightarrow\infty]{}{\text{limsup}}h(Q_{n_k}, x)\leq h(Q, x)$ for any $x\in R^d$, we have,
\begin{eqnarray}
\stackrel[k\rightarrow\infty]{}{\text{liminf}}(\phi_{n_k}(x_0)) &\geq& \frac{\lambda- \int \text{log}(m(x)+1)Q(dx) -\text{max}(M_*,0)}{1-h(Q,x_0)}>-\infty. \nonumber
\end{eqnarray}
Hence, for k large enough,
\begin{equation}\label{infimumofphi2}
\stackrel[l\geq k]{}{\text{inf}}\phi_{n_l}(x_0)>-\infty, \ \forall x_0\in \text{interior}(\text{csupp}(Q)).
\end{equation}
Again, we can deduce from (\ref{infimumofphi2}) and the boundedness of $M_{n_k}$ that there exist constants $a$ and $b>0$ such that,
\begin{equation}\label{commonbound2}
\phi_{n_k}(x)\leq a-b||x||, \ \forall k \ \text{sufficiently large}, \ x\in R^d.
\end{equation}
Similar as before we conclude that there exist $\phi_*\in\Phi^d$ and a subsequence $\{\phi_{n_{k_l}}\}$ such that $\text{interior}(\text{csupp}(Q))\subseteq \text{dom}(\phi_*)\subseteq \text{csupp}(Q)$ and,
\begin{eqnarray}
  \stackrel[l\rightarrow\infty,x\rightarrow y]{}{\text{limsup}}\phi_{n_{k_l}}(x) &\leq& \phi_*(y)\leq a-b||y||, \  \forall y\in R^d,  \nonumber\\
  \stackrel[l\rightarrow\infty,x\rightarrow y]{}{\text{lim}}\phi_{n_{k_l}}(x) &=& \phi_*(y)>-\infty, \ \forall y\in \text{interior}(\text{csupp}(Q)).  \nonumber
\end{eqnarray}
Then $\int e^{\phi_*(x)}dx=1$ by dominated convergence, which implies that $\phi_*\in\tilde{\Phi}^d$.

By Skorohod's theorem, there exist a probability space $(\Omega, \mathcal{F}, P)$ and random variables $X_{n_{k_l}}\sim Q_{n_{k_l}}$, $X\sim Q$, such that $\stackrel[l\rightarrow\infty]{}{\text{lim}}X_n=X$ almost surely. Let $H_{n_{k_l}}= \int \text{log}(m(x)+1)Q (dx)+a-b||X_{n_{k_l}}||-\text{log}\{(1-p_{n_{k_l}})f_0(X_{n_{k_l}})+p_{n_{k_l}}\text{exp}(\phi_{n_{k_l}}(X_{n_{k_l}}))\}$.  By Fatou's Lemma, we have,
 \begin{eqnarray}
   \lambda &=& \stackrel[l\rightarrow\infty]{}{\text{lim}}L(p_{n_{k_l}}, \phi_{n_{k_l}}, Q_{n_{k_l}})=\stackrel[l\rightarrow\infty]{}{\text{lim}}\int \text{log}((1-p_{n_{k_l}})f_0+p_{n_{k_l}}e^{\phi_{n_{k_l}}})dQ_{n_{k_l}} \nonumber \\
   &=& \stackrel[l\rightarrow\infty]{}{\text{lim}}\{ \int \text{log}(m(x)+1)Q (dx)+\int( a-b||x||)Q_{n_{k_l}}(dx)-E(H_{n_{k_l}})\} \nonumber\\
   &=&  \int \text{log}(m(x)+1)Q(dx)+a-b\int ||x||Q(dx)-\stackrel[l\rightarrow\infty]{}{\text{liminf}}E(H_{n_{k_l}}) \nonumber\\
   &\leq&  \int \text{log}(m(x)+1)Q(dx)+a-b\int ||x||Q(dx)-E(\stackrel[l\rightarrow\infty]{}{\text{liminf}}(H_{n_{k_l}}))  \nonumber\\
   &\leq& E\{ \stackrel[l\rightarrow\infty]{}{\text{limsup}}\text{log}((1-p_{n_{k_l}})f_0(X_{n_{k_l}})+p_{n_{k_l}}\text{exp}(\phi_{n_{k_l}}(X_{n_{k_l}})))\} \nonumber\\
   &\leq& E(\text{log}((1-p_*)f_0(X)+p_*\text{exp}(\phi_*(X))))\nonumber\\
   &\leq& L(Q).\nonumber
\end{eqnarray}

In order to show that $\lambda\geq L(Q)$, we use the approximations $\phi^*\leq{\phi^*}^{(\epsilon)}\leq{\phi^*}^{(1)}$, $0<\epsilon\leq 1$ from Lemma 4.4 of \cite{schuhmacher2011multivariate}, since ${\phi^*}^{(\epsilon)}\in\Phi^d$ is Lipschitz continuous, one can show that $\frac{|{\phi^*}^{(\epsilon)}|}{1+||x||}$ is bounded, and hence by Lemma 4.6 of \cite{schuhmacher2011multivariate}, we have,

\begin{eqnarray}
   \lambda &=& \stackrel[k\rightarrow\infty]{}{\text{lim}}L(p_{n_k},\phi_{n_k}, Q_{n_k})  \nonumber\\
   &\geq& \stackrel[k\rightarrow\infty]{}{\text{lim}}L(p^*, {\phi^*}^{(\epsilon)}-\text{log}(\int e^{{\phi^*}^{(\epsilon)}(x)} dx), Q_{n_k})  \nonumber\\
   &=&  L(p^*, {\phi^*}^{(\epsilon)}-\text{log}(\int e^{{\phi^*}^{(\epsilon)}(x)} dx), Q) \nonumber\\
   &=& \int \text{log}\{(1-p^*)f_0\int e^{{\phi^*}^{(\epsilon)}(x)}dx+p^*e^{{\phi^*}^{(\epsilon)}}\}dQ-\text{log}(\int e^{{\phi^*}^{(\epsilon)}(x)}dx) \nonumber\\
   &\rightarrow& \int \text{log}((1-p^*)f_0+p^*e^{\phi^*})dQ=L(p^*,\phi^*,Q), \ \text{as} \ \epsilon\rightarrow 0.  \nonumber
\end{eqnarray}
The last step above is by applying dominated convergence on $e^{{\phi^*}^{(\epsilon)}}$ and monotone convergence on $(1-p^*)f_0\int e^{{\phi^*}^{(1)}(x)}dx+p^*e^{{\phi^*}^{(1)}}-(1-p^*)f_0\int e^{{\phi^*}^{(\epsilon)}(x)}dx-p^*e^{{\phi^*}^{(\epsilon)}}$. Thus we have shown that $\lambda=L(Q)$, and $(p^*,\phi^*)=(p_*,\phi_*)$ is the unique maximizer.

With exactly the same argument, we can show that $\stackrel[n\rightarrow\infty]{}{\text{liminf}}L(Q_n)=L(Q)$ as well, and hence $L(Q_n)\rightarrow L(Q)$, as $n\rightarrow\infty$.

Also, if we let $f^*=\text{exp}\circ \phi^*$, $f_n=\text{exp}\circ \phi_n$, we have shown that,
\begin{eqnarray}
  \stackrel[l\rightarrow\infty]{}{\text{lim}}p_{n_{k_l}}&=& p^*, \nonumber\\
  \stackrel[l\rightarrow\infty,x\rightarrow y]{}{\text{limsup}}f_{n_{k_l}}(x) &\leq& f^*(y), \  \forall y\in\partial\{f^*>0\},  \nonumber\\
  \stackrel[l\rightarrow\infty,x\rightarrow y]{}{\text{lim}}f_{n_{k_l}}(x) &=& f^*(y), \ \forall y\in R^d\setminus \partial\{f^*>0\}.  \nonumber
\end{eqnarray}
In particular, $\{f_{n_{k_l}}\}$ converges to $f^*$ almost everywhere w.r.t. Lebesgue measure and hence $\int|f_{n_{k_l}}(x)-f^*(x)|dx\rightarrow 0$, as $l\rightarrow\infty$ by dominated convergence. Our proof actually shows that for any subsequence of $\{Q_n\}$, we can further find a subsequence with the above convergence properties. That means the original sequence must satisfy those properties as well, otherwise we would arrive at contradictions and that completes the proof.

\end{proof}\hfill\\

\subsection{More Simulation Result}

\begin{table}[H]

\caption{Bias(MSE) of estimates of  $p/\mu$ and mean of the classification error for model 1 when $n=250$.}
\centering
\tabcolsep=0.08cm
\renewcommand*{\arraystretch}{0.8}
\begin{tabular}{c|cccccc}
  \hline
      &  EM\_logconcave &  Patra & Bordes &  Song EM &   Song max $\pi$  &   Xiang \\
   \hline
  \multicolumn{1}{c}{$p=0.2$} &   &   &   &  & & \\
  \hline
  \multicolumn{1}{c|}{$p$} &  0.008(0.0019) &  -0.001(0.0021) & 0.018(0.0046)  & -0.071(0.0057) & 0.106(0.0160) & 0.021(0.0056)\\
   \multicolumn{1}{c|}{$\mu$}  & 0.057(0.0738)  & -0.396(0.3286)  & -0.166(0.2109) & -0.108(0.2243) & -0.846(0.9437) & 0.243(0.1864)\\
   \multicolumn{1}{c|}{Cla\_error}  &  0.1029 & 0.1094  & 0.1097  & 0.1138 & 0.1104 & 0.1058\\
   \hline
  \multicolumn{1}{c}{$p=0.5$} &   &   &   &  & & \\
  \hline
  \multicolumn{1}{c|}{$p$} & 0.000(0.0023)  & -0.041(0.0036)  & 0.005(0.0025)  & -0.130(0.0185) & 0.100(0.0153) & 0.014(0.0026)\\
   \multicolumn{1}{c|}{$\mu$}  & 0.017(0.0225)  & 0.023(0.0167)  & -0.021(0.0198) & 0.143(0.0393) & -0.344(0.1635) & 0.032(0.0208)\\
   \multicolumn{1}{c|}{Cla\_error}  & 0.1151  & 0.1259  & 0.1232  & 0.1379 & 0.1248 & 0.1138 \\
   \hline
  \multicolumn{1}{c}{$p=0.8$} &   &   &   &  & & \\
  \hline
  \multicolumn{1}{c|}{$p$} & -0.001(0.0011)  & -0.070(0.0057)  & -0.001(0.0014)  & -0.104(0.0123) & 0.056(0.0040) & 0.016(0.0014)\\
   \multicolumn{1}{c|}{$\mu$}  & 0.003(0.0072)  & 0.059(0.0102)  & -0.001(0.0085) & 0.097(0.0158) & -0.147(0.0323) & -0.008(0.0079)\\
   \multicolumn{1}{c|}{Cla\_error}  & 0.0670  & 0.0781  & 0.0722  & 0.0835 & 0.0752 & 0.0703 \\
   \hline
\end{tabular}
\end{table}

\begin{table}[H]
\caption{Bias(MSE) of estimates of  $p/\mu$ and mean of the classification error for model 1 when $n=500$.}
\centering
\tabcolsep=0.08cm
\renewcommand*{\arraystretch}{0.8}
\begin{tabular}{c|cccccc}
   \hline
      &  EM\_logconcave &  Patra & Bordes &  Song EM &   Song max $\pi$  &   Xiang \\
   \hline
  \multicolumn{1}{c}{$p=0.2$} &   &   &   &  & & \\
  \hline
  \multicolumn{1}{c|}{$p$} & 0.000(0.0008)  & -0.008(0.0014)  & 0.003(0.0021)  & -0.077(0.0063) & 0.086(0.0102) & 0.011(0.0013)\\
   \multicolumn{1}{c|}{$\mu$}  & 0.059(0.0348)  & -0.258(0.1430)  & -0.054(0.1013) & 0.001(0.0734) & -0.738(0.6525) & 0.175(0.0843)\\
   \multicolumn{1}{c|}{Cla\_error}  & 0.0972  & 0.1070  & 0.1060  & 0.1106 & 0.1051 & 0.0990 \\
   \hline
  \multicolumn{1}{c}{$p=0.5$} &   &   &   &  & & \\
  \hline
  \multicolumn{1}{c|}{$p$} &  -0.003(0.0009) & -0.031(0.0021)  & 0.000(0.0012)  & -0.132(0.0181) & 0.107(0.0158) & 0.011(0.0011)\\
   \multicolumn{1}{c|}{$\mu$}  & 0.019(0.0097)  & 0.035(0.0109)  & -0.003(0.0098) & 0.169(0.0363) & -0.346(0.1605) & 0.020(0.0097)\\
   \multicolumn{1}{c|}{Cla\_error}  & 0.1111  & 0.1239  & 0.1209  & 0.1359 & 0.1226 & 0.1120\\
   \hline
  \multicolumn{1}{c}{$p=0.8$} &   &   &   &  & & \\
  \hline
  \multicolumn{1}{c|}{$p$} & 0.003(0.0006)  & -0.053(0.0033)  & 0.001(0.0007)  & -0.104(0.0117) & 0.056(0.0040) & 0.014(0.0006)\\
   \multicolumn{1}{c|}{$\mu$}  &  -0.001(0.0032) &  0.065(0.0073) & 0.000(0.0041) & 0.110(0.0155) & -0.121(0.0220) & -0.007(0.0040)\\
   \multicolumn{1}{c|}{Cla\_error}  & 0.0644  & 0.0758  & 0.0693  & 0.0822 & 0.0711 & 0.0685\\
   \hline
\end{tabular}
\end{table}

\begin{table}[H]
\caption{Bias(MSE) of estimates of  $p/\mu$ and mean of the classification aerror for model 2 when $n=250$.}
\centering
\tabcolsep=0.08cm
\renewcommand*{\arraystretch}{0.8}
\begin{tabular}{c|cccccc}
 \hline
      &  EM\_logconcave &  Patra & Bordes &  Song EM &   Song max $\pi$  &   Xiang \\
   \hline
  \multicolumn{1}{c}{$p=0.2$} &   &   &   &  & & \\
  \hline
  \multicolumn{1}{c|}{$p$} & 0.004(0.0051)  & -0.021(0.0033)  & -0.006(0.0036)  & -0.156(0.0248) & 0.371(0.1443) & 0.056(0.0087)\\
   \multicolumn{1}{c|}{$\mu$}  & -0.022(0.0038)  & 0.061(0.0073)  & -0.019(0.0029) & 0.013(0.0032) & 0.197(0.0401) & -0.011(0.0026)\\
   \multicolumn{1}{c|}{Cla\_error}  & 0.1368  & 0.1554  & 0.1568  & 0.1746 & 0.1858 & 0.1437\\
   \hline
  \multicolumn{1}{c}{$p=0.5$} &   &   &   &  & & \\
  \hline
  \multicolumn{1}{c|}{$p$} & 0.004(0.0043)  & -0.064(0.0071)  & -0.037(0.0041)  & -0.300(0.0916) & 0.230(0.0576) & -0.013(0.0041)\\
   \multicolumn{1}{c|}{$\mu$}  & -0.005(0.0007)  & -0.004(0.0004)  & -0.032(0.0013) & -0.034(0.0013) & 0.080(0.0070) & -0.014(0.0011)\\
   \multicolumn{1}{c|}{Cla\_error}  & 0.1678  & 0.2110  & 0.2031  & 0.2778 & 0.1964 & 0.1764\\
   \hline
  \multicolumn{1}{c}{$p=0.8$} &   &   &   &  & & \\
  \hline
  \multicolumn{1}{c|}{$p$} & 0.009(0.0019)  & -0.105(0.0124)  & -0.065(0.0057)  & -0.312(0.1001) & 0.081(0.0078) & -0.049(0.0056)\\
   \multicolumn{1}{c|}{$\mu$}  & 0.000(0.0002)  & -0.020(0.0005)  & -0.033(0.0012) & -0.039(0.0016) & 0.010(0.0006) & -0.018(0.0010)\\
   \multicolumn{1}{c|}{Cla\_error}  & 0.1030  & 0.1384  & 0.1266  & 0.2137 & 0.1124 & 0.1140 \\
   \hline
\end{tabular}
\end{table}

\begin{table}[H]
\caption{Bias(MSE) of estimates of  $p/\mu$ and mean of the classification error for model 2 when $n=500$.}
\centering
\tabcolsep=0.08cm
\renewcommand*{\arraystretch}{0.8}
\begin{tabular}{c|cccccc}
   \hline
      &  EM\_logconcave &  Patra & Bordes &  Song EM &   Song max $\pi$  &   Xiang \\
   \hline
  \multicolumn{1}{c}{$p=0.2$} &   &   &   &  & & \\
  \hline
  \multicolumn{1}{c|}{$p$} & -0.001(0.0028)  &  -0.021(0.0020) & -0.007(0.0022)  & -0.154(0.0238) & 0.379(0.1496) & 0.035(0.0047)\\
   \multicolumn{1}{c|}{$\mu$}  & -0.019(0.0024)  & 0.041(0.0030) & -0.025(0.0022) & -0.006(0.0008) & 0.197(0.0393) & -0.009(0.0018) \\
   \multicolumn{1}{c|}{Cla\_error}  &  0.1294 & 0.1544  & 0.1520  & 0.1697 & 0.1862 & 0.1369\\
   \hline
  \multicolumn{1}{c}{$p=0.5$} &   &   &   &  & & \\
  \hline
  \multicolumn{1}{c|}{$p$} &  0.002(0.0022) & -0.053(0.0045)  & -0.039(0.0032)  & -0.292(0.0860) & 0.234(0.0583) & -0.034(0.0035)\\
   \multicolumn{1}{c|}{$\mu$}  & -0.004(0.0003)  & -0.008(0.0002)  & -0.033(0.0013) & -0.037(0.0014) & 0.080(0.0069) & -0.014(0.0009)\\
   \multicolumn{1}{c|}{Cla\_error}  & 0.1638  & 0.2031  & 0.2011  & 0.2723 & 0.1940 & 0.1753\\
   \hline
\multicolumn{1}{c}{$p=0.8$} &   &   &   &  & & \\
  \hline
  \multicolumn{1}{c|}{$p$} & 0.003(0.0010)  & -0.086(0.0081)  & -0.070(0.0058)  & -0.312(0.0990) & 0.097(0.0102) & -0.048(0.0038)\\
   \multicolumn{1}{c|}{$\mu$}  & -0.001(0.0001)  & -0.020(0.0005)  & -0.034(0.0012) & -0.040(0.0016) & 0.024(0.0007) & -0.022(0.0010)\\
   \multicolumn{1}{c|}{Cla\_error}  & 0.1001  & 0.1307  & 0.1263  & 0.2119 & 0.1129 & 0.1139 \\
   \hline
\end{tabular}
\end{table}

\begin{table}[H]
\caption{Bias(MSE) of estimates of  $p/\mu$ and mean of the classification error for model 3 when $n=250$.}
\centering
\tabcolsep=0.08cm
\renewcommand*{\arraystretch}{0.8}
\begin{tabular}{c|cccccc}
   \hline
      &  EM\_logconcave &  Patra & Bordes &  Song EM &   Song max $\pi$  &   Xiang \\
   \hline
  \multicolumn{1}{c}{$p=0.2$} &   &   &   &  & & \\
  \hline
  \multicolumn{1}{c|}{$p$} & 0.005(0.0011)  & 0.009(0.0021)  & NA  & -0.050(0.0034) & 0.431(0.1889) & 0.048(0.0042)\\
   \multicolumn{1}{c|}{$\mu$}  & 0.026(0.0400)  & -0.041(0.0581)  & NA & 0.048(0.0591) & -1.115(1.2677) & -0.139(0.0493)\\
   \multicolumn{1}{c|}{Cla\_error}  & 0.0737  & 0.0869  & NA  & 0.0895 & 0.1718 & 0.0842\\
   \hline
  \multicolumn{1}{c}{$p=0.5$} &   &   &   &  & & \\
  \hline
  \multicolumn{1}{c|}{$p$} & 0.002(0.0013)  & -0.019(0.0019)  & NA  & -0.069(0.0065) & 0.269(0.0742) & 0.081(0.0103)\\
   \multicolumn{1}{c|}{$\mu$}  & -0.001(0.0096)  & -0.001(0.0126)  & NA & 0.013(0.0120) & -0.495(0.2618) & -0.174(0.0590)\\
   \multicolumn{1}{c|}{Cla\_error}  & 0.0623  & 0.0806  & NA  & 0.0839 & 0.1271 & 0.0860\\
   \hline
  \multicolumn{1}{c}{$p=0.8$} &   &   &   &  & & \\
  \hline
  \multicolumn{1}{c|}{$p$} & 0.003(0.0007)  & -0.046(0.0029)  & NA  & -0.225(0.0017) & 0.107(0.0122) & 0.087(0.0096)\\
   \multicolumn{1}{c|}{$\mu$}  & 0.001(0.0052)  &  0.003(0.0061) & NA & -0.004(0.0057) & -0.157(0.0316) & -0.150(0.0446)\\
   \multicolumn{1}{c|}{Cla\_error}  & 0.0274  & 0.0351  & NA  & 0.0354 & 0.0589 & 0.0734\\
   \hline
\end{tabular}
\end{table}

\begin{table}[H]
\caption{Bias(MSE) of estimates of  $p/\mu$ and mean of the classification error for model 3 when $n=500$.}
\centering
\tabcolsep=0.08cm
\renewcommand*{\arraystretch}{0.8}
\begin{tabular}{c|cccccc}
   \hline
      &  EM\_logconcave &  Patra & Bordes &  Song EM &   Song max $\pi$  &   Xiang \\
   \hline
  \multicolumn{1}{c}{$p=0.2$} &   &   &   &  & & \\
  \hline
  \multicolumn{1}{c|}{$p$} & 0.004(0.0005)  & 0.004(0.0011)  & NA  & -0.055(0.0034) & 0.415(0.1746) & 0.038(0.0024)\\
   \multicolumn{1}{c|}{$\mu$}  & 0.014(0.0204)  & -0.039(0.0316)  & NA & 0.047(0.0315) & -1.119(1.2657) & -0.131(0.0357)\\
   \multicolumn{1}{c|}{Cla\_error}  & 0.0722  & 0.0862  & NA  & 0.0855 & 0.1610 & 0.0819\\
   \hline
  \multicolumn{1}{c}{$p=0.5$} &   &   &   &  & & \\
  \hline
  \multicolumn{1}{c|}{$p$} & 0.001(0.0005)  & -0.016(0.0010)  & NA  & -0.070(0.0057) & 0.260(0.0692) & 0.060(0.0059)\\
   \multicolumn{1}{c|}{$\mu$}  & 0.004(0.0047)  & -0.007(0.0061)  & NA & 0.017(0.0064) & -0.489(0.2475) & -0.126(0.0338)\\
   \multicolumn{1}{c|}{Cla\_error}  & 0.0604  & 0.0787  & NA  & 0.0811 & 0.1189 & 0.0790\\
   \hline
  \multicolumn{1}{c}{$p=0.8$} &   &   &   &  & & \\
  \hline
  \multicolumn{1}{c|}{$p$} & 0.002(0.0003)  & -0.036(0.0017)  & NA  & -0.029(0.0013) & 0.106(0.0115) & 0.080(0.0078)\\
   \multicolumn{1}{c|}{$\mu$}  & 0.001(0.0027)  & 0.001(0.0026)  & NA & -0.002(0.0030) & -0.159(0.0294) & -0.117(0.0284)\\
   \multicolumn{1}{c|}{Cla\_error}  & 0.0270  & 0.0334  & NA  & 0.0341 & 0.0557 & 0.0677\\
   \hline
\end{tabular}
\end{table}

\begin{table}[H]
\caption{Bias(MSE) of estimates of  $p/\mu$ and mean of the classification error for model 4 when $n=250$.}
\centering
\tabcolsep=0.08cm
\renewcommand*{\arraystretch}{0.8}
\begin{tabular}{c|cccccc}
   \hline
      &  EM\_logconcave &  Patra & Bordes &  Song EM &   Song max $\pi$  &   Xiang \\
   \hline
  \multicolumn{1}{c}{$p=0.2$} &   &   &  & & & \\
  \hline
  \multicolumn{1}{c|}{$p$} & -0.001(0.0009)  & 0.019(0.0021)  & NA  & 0.019(0.0013) & 0.129(0.0198) & 0.113(0.0179)\\
   \multicolumn{1}{c|}{$\mu$}  & 0.016(0.1454)  & -0.574(0.5652)  & NA & -0.569(0.5268) & -1.296(2.0126) & -0.728(1.0325)\\
   \multicolumn{1}{c|}{Cla\_error}  & 0.0128  & 0.0168  & NA  & 0.0168 & 0.0247 & 0.0438\\
   \hline
  \multicolumn{1}{c}{$p=0.5$} &   &   &   &  & & \\
  \hline
  \multicolumn{1}{c|}{$p$} &  0.003(0.0009) & -0.014(0.0014)  & NA  & 0.022(0.0014) & 0.085(0.0089) & 0.086(0.0101)\\
   \multicolumn{1}{c|}{$\mu$}  & 0.025(0.0493)  & -0.192(0.0898)  & NA & -0.200(0.0878) & -0.379(0.2136) & -0.605(0.6314)\\
   \multicolumn{1}{c|}{Cla\_error}  & 0.0096 & 0.0184  & NA  & 0.0188 & 0.0204 & 0.0378\\
   \hline
  \multicolumn{1}{c}{$p=0.8$} &   &   &   &  & & \\
  \hline
  \multicolumn{1}{c|}{$p$} & 0.000(0.0006)  & -0.039(0.0022)  & NA  & 0.013(0.0007) & 0.044(0.0026) & 0.077(0.0076)\\
   \multicolumn{1}{c|}{$\mu$}  & 0.009(0.0279)  & -0.024(0.0247)  & NA & -0.073(0.0332) & -0.157(0.0572) & -0.621(0.4861)\\
   \multicolumn{1}{c|}{Cla\_error}  & 0.0044  & 0.0093  & NA  & 0.0115 & 0.0149 & 0.0468\\
   \hline
\end{tabular}
\end{table}

\begin{table}[H]
\caption{Bias(MSE) of estimates of  $p/\mu$ and mean of the classification error for model 4 when $n=500$.}
\centering
\tabcolsep=0.08cm
\renewcommand*{\arraystretch}{0.8}
\begin{tabular}{c|cccccc}
   \hline
      &  EM\_logconcave &  Patra & Bordes &  Song EM &   Song max $\pi$  &   Xiang \\
   \hline
  \multicolumn{1}{c}{$p=0.2$} &   &   &   &  & & \\
  \hline
  \multicolumn{1}{c|}{$p$} & -0.002(0.0003)  &  0.009(0.0009) & NA  & 0.010(0.0005) & 0.108(0.0141) & 0.074(0.0073)\\
   \multicolumn{1}{c|}{$\mu$}  & -0.008(0.0721)  & -0.410(0.2935)  & NA & -0.404(0.2668) & -1.109(1.4363) & -0.803(1.1190)\\
   \multicolumn{1}{c|}{Cla\_error}  &  0.0116 & 0.0151  & NA  & 0.0147 & 0.0201 & 0.0271\\
   \hline
  \multicolumn{1}{c}{$p=0.5$} &   &   &   &  & & \\
  \hline
  \multicolumn{1}{c|}{$p$} & 0.000(0.0005)  & -0.011(0.0008)  & NA  & 0.017(0.0009) & 0.074(0.0066) & 0.069(0.0062)\\
   \multicolumn{1}{c|}{$\mu$}  & -0.011(0.0244)  & -0.169(0.0525)  & NA & -0.220(0.0720) & -0.374(0.1801) & -0.607(0.6089)\\
   \multicolumn{1}{c|}{Cla\_error}  &  0.0095 & 0.0167  & NA  & 0.0178 & 0.0178 & 0.0288\\
   \hline
  \multicolumn{1}{c}{$p=0.8$} &   &   &   &  & & \\
  \hline
  \multicolumn{1}{c|}{$p$} &  0.002(0.0004) & -0.031(0.0014)  & NA  & 0.011(0.0005) & 0.042(0.0023) & 0.068(0.0054)\\
   \multicolumn{1}{c|}{$\mu$}  & -0.007(0.0137)  & -0.034(0.0151)  & NA & -0.072(0.0190) & -0.159(0.0447) & -0.699(0.5384)\\
   \multicolumn{1}{c|}{Cla\_error}  & 0.0048  & 0.0082  & NA  & 0.0103& 0.0126 & 0.0360\\
  \hline
\end{tabular}
\end{table}

\begin{table}[H]
\caption{Bias(MSE) of estimates of  $p/\mu$ and mean of the classification error for model 5 when $n=250$.}
\centering
\tabcolsep=0.08cm
\renewcommand*{\arraystretch}{0.8}
\begin{tabular}{c|cccccc}
  \hline
      &  EM\_logconcave &  Patra & Bordes &  Song EM &   Song max $\pi$  &   Xiang \\
   \hline
  \multicolumn{1}{c}{$p=0.2$} &   &   &   &  & & \\
  \hline
  \multicolumn{1}{c|}{$p$} & -0.001(0.0007)  &  0.020(0.0022) & NA  & 0.030(0.0018) & 0.142(0.0240) & 0.047(0.0036)\\
   \multicolumn{1}{c|}{$\mu$}  & -0.026(0.0943)  & -0.679(0.6177)  & NA & -0.716(0.6592) & -1.449(2.3985) & -0.895(0.9545)\\
   \multicolumn{1}{c|}{Cla\_error}  & 0.0017  &  0.0090 & NA  & 0.0087 & 0.0185 & 0.0127\\
   \hline
  \multicolumn{1}{c}{$p=0.5$} &   &   &   &  & & \\
  \hline
  \multicolumn{1}{c|}{$p$} & 0.001(0.0010)  & -0.015(0.0014)  & NA  & 0.031(0.0019) & 0.116(0.0162) & 0.074(0.0075)\\
   \multicolumn{1}{c|}{$\mu$}  & 0.009(0.0288)  & -0.208(0.0734)  & NA & -0.267(0.1017) & -0.613(0.4915) & -0.680(0.5574)\\
   \multicolumn{1}{c|}{Cla\_error}  & 0.0011  & 0.0114  & NA  & 0.0136 & 0.0202 & 0.0248\\
   \hline
  \multicolumn{1}{c}{$p=0.8$} &   &   &   &  & & \\
  \hline
  \multicolumn{1}{c|}{$p$} &  0.000(0.0006) & -0.042(0.0025)  & NA  & 0.015(0.0008) & 0.069(0.0056) & 0.089(0.0096)\\
   \multicolumn{1}{c|}{$\mu$}  & 0.006(0.0216)  & -0.036(0.0232)  & NA & -0.089(0.0305) & -0.296(0.1188) & -0.488(0.3336)\\
   \multicolumn{1}{c|}{Cla\_error}  & 0.0004  & 0.0037  & NA  & 0.0055 & 0.0188 & 0.0544\\
   \hline
\end{tabular}
\end{table}

\begin{table}[H]
\caption{Bias(MSE) of estimates of  $p/\mu$ and mean of the classification error for model 5 when $n=500$.}
\centering
\tabcolsep=0.08cm
\renewcommand*{\arraystretch}{0.8}
\begin{tabular}{c|cccccc}
  \hline
      &  EM\_logconcave &  Patra & Bordes &  Song EM &   Song max $\pi$  &   Xiang \\
   \hline
  \multicolumn{1}{c}{$p=0.2$} &   &   &   &  & & \\
  \hline
  \multicolumn{1}{c|}{$p$} & -0.001(0.0003)  & 0.012(0.0011)  & NA  & 0.023(0.0009) & 0.119(0.0176) & 0.042(0.0025)\\
   \multicolumn{1}{c|}{$\mu$}  & -0.007(0.0438)  & -0.500(0.3260)  & NA & -0.553(0.3655) & 1.237(1.7273) & -0.928(0.9531)\\
   \multicolumn{1}{c|}{Cla\_error}  & 0.0014  & 0.0058  & NA  & 0.0061 & 0.0132 & 0.0107\\
   \hline
  \multicolumn{1}{c}{$p=0.5$} &   &   &   &  & & \\
  \hline
  \multicolumn{1}{c|}{$p$} &  -0.001(0.0006) & -0.010(0.0007)  & NA  & 0.025(0.0012) & 0.120(0.0179) & 0.077(0.0073)\\
   \multicolumn{1}{c|}{$\mu$}  & 0.014(0.0170)  & -0.194(0.0565)  & NA & -0.233(0.0722) & -0.630(0.4822) & -0.726(0.5840)\\
   \multicolumn{1}{c|}{Cla\_error}  & 0.0008  &  0.0094 & NA  & 0.0111 & 0.0188 & 0.0240\\
   \hline
  \multicolumn{1}{c}{$p=0.8$} &   &   &   &  & & \\
  \hline
  \multicolumn{1}{c|}{$p$} & 0.001(0.0003)  & -0.031(0.0013)  & NA  & 0.013(0.0004) & 0.078(0.0067) & 0.088(0.0087)\\
   \multicolumn{1}{c|}{$\mu$}  & 0.006(0.0094)  & -0.020(0.0105)  & NA & -0.065(0.0146) & -0.324(0.1188) & -0.506(0.3185)\\
   \multicolumn{1}{c|}{Cla\_error}  & 0.0003  & 0.0026  & NA  & 0.0039 & 0.0199 & 0.0501\\
   \hline
\end{tabular}
\end{table}

\begin{table}[H]
\caption{Bias(MSE) of estimates of  $p/\mu$ and mean of the classification error for model 6 when $n=250$.}
\centering
\tabcolsep=0.08cm
\renewcommand*{\arraystretch}{0.8}
\begin{tabular}{c|cccccc}
   \hline
      &  EM\_logconcave &  Patra & Bordes &  Song EM &   Song max $\pi$  &   Xiang \\
   \hline
  \multicolumn{1}{c}{$p=0.2$} &   &   &   &  & & \\
  \hline
  \multicolumn{1}{c|}{$p$} & -0.006(0.0015)  & 0.003(0.0021)  & 0.020(0.0036)  & -0.012(0.0009) & 0.109(0.0147) & 0.027(0.0023)\\
   \multicolumn{1}{c|}{$\mu$}  & 0.127(0.1230)  & -0.200(0.1172)  & -0.685(0.8830) & -0.118(0.0777) & -0.654(0.5415) & -0.067(0.0877)\\
   \multicolumn{1}{c|}{Cla\_error}  & 0.0464  & 0.0468  & 0.1738  & 0.0470 & 0.0509 & 0.0473\\
   \hline
  \multicolumn{1}{c}{$p=0.5$} &   &   &   &  & & \\
  \hline
  \multicolumn{1}{c|}{$p$} & -0.015(0.0028)  & -0.044(0.0037)  & 0.000(0.0024)  & -0.045(0.0031) & 0.057(0.0054) & 0.031(0.0061)\\
   \multicolumn{1}{c|}{$\mu$}  & 0.073(0.0525)  & 0.123(0.0325)  & -0.009(0.0298) & 0.127(0.0329) & -0.099(0.0405) & -0.068(0.0676)\\
   \multicolumn{1}{c|}{Cla\_error}  & 0.0688  & 0.0682  & 0.0676  & 0.0688 & 0.0683 & 0.0738\\
   \hline
  \multicolumn{1}{c}{$p=0.8$} &   &   &   &  & & \\
  \hline
  \multicolumn{1}{c|}{$p$} & 0.006(0.0016)  & -0.077(0.0067)  & -0.003(0.0016)  & -0.059(0.0044) & 0.011(0.0012) & 0.006(0.0013)\\
   \multicolumn{1}{c|}{$\mu$}  & -0.024(0.0205)  & 0.173(0.0369)  & 0.004(0.0104) & 0.155(0.0320) & 0.032(0.0125) & -0.002(0.0105)\\
   \multicolumn{1}{c|}{Cla\_error}  & 0.0591  & 0.0660  & 0.0595  & 0.0647 & 0.0588 & 0.0572\\
   \hline
\end{tabular}
\end{table}

\begin{table}[H]
\caption{Bias(MSE) of estimates of  $p/\mu$ and mean of the classification error for model 6 when $n=500$.}
\centering
\tabcolsep=0.08cm
\renewcommand*{\arraystretch}{0.8}
\begin{tabular}{c|cccccc}
  \hline
      &  EM\_logconcave &  Patra & Bordes &  Song EM &   Song max $\pi$  &   Xiang \\
   \hline
  \multicolumn{1}{c}{$p=0.2$} &   &   &   &  & & \\
  \hline
  \multicolumn{1}{c|}{$p$} & -0.008(0.0006)  & -0.001(0.0008)  & 0.016(0.0021)  & -0.017(0.0007) & 0.095(0.0119) & 0.011(0.0007)\\
   \multicolumn{1}{c|}{$\mu$}  & 0.132(0.0565)  & -0.066(0.0452)  & -0.161(0.1613) & 0.000(0.0312) & -0.538(0.3834) & -0.025(0.0370)\\
   \multicolumn{1}{c|}{Cla\_error}  & 0.0435  & 0.0457  & 0.0451  & 0.0450 & 0.0479 & 0.0439\\
   \hline
  \multicolumn{1}{c}{$p=0.5$} &   &   &   &  & & \\
  \hline
  \multicolumn{1}{c|}{$p$} & -0.011(0.0014)  & -0.033(0.0019)  & -0.001(0.0010)  & -0.049(0.0030) & 0.043(0.0032) & 0.011(0.0017)\\
   \multicolumn{1}{c|}{$\mu$}  & 0.069(0.0304)  & 0.137(0.0272)  & -0.004(0.0109) & 0.161(0.0339) & -0.038(0.0183) & -0.023(0.0203)\\
   \multicolumn{1}{c|}{Cla\_error}  & 0.0655  & 0.0676  & 0.0666  & 0.0684 & 0.0665 & 0.0664\\
   \hline
  \multicolumn{1}{c}{$p=0.8$} &   &   &   &  & & \\
  \hline
  \multicolumn{1}{c|}{$p$} & 0.004(0.0005)  & -0.065(0.0047)  & -0.002(0.0008)  & -0.062(0.0043) & 0.010(0.0011) & 0.002(0.0007)\\
   \multicolumn{1}{c|}{$\mu$}  & -0.021(0.0069)  & 0.169(0.0325)  & -0.002(0.0054) & 0.165(0.0311) & 0.043(0.0102) & 0.002(0.0060)\\
   \multicolumn{1}{c|}{Cla\_error}  & 0.0541  & 0.0646  & 0.0576  & 0.0642 & 0.0574 & 0.0546\\
   \hline
\end{tabular}
\end{table}\hfill

\section*{Acknowledgment}
Yao's research is supported by NSF grant DMS-1461677 and Department of Energy with the award
No: 10006272.\\



\bibliography{bibfile,mixture-ref}
\bibliographystyle{apalike}

\end{document}